\documentclass{article}
\usepackage{jheppub}

 \normalsize

\def\lsim{\raise0.3ex\hbox{$\;<$\kern-0.75em\raise-1.1ex\hbox{$\sim\;$}}}
\def\gsim{\raise0.3ex\hbox{$\;>$\kern-0.75em\raise-1.1ex\hbox{$\sim\;$}}}

\usepackage{graphics}

\newcommand{\be}{\begin{eqnarray}}
\newcommand{\ee}{\end{eqnarray}}

\def\bea{\begin{eqnarray}}
\def\eea{\end{eqnarray}}

\usepackage{epsfig}
 \normalsize

\begin{document}

\title{Enhancement of $H \to \gamma \gamma$ in $SU(5)$ model with $45_H$-plet}
\author{S. Khalil$^{1,2,3}$ and S. Salem$^{1,4}$}
\vspace*{0.2cm}
\affiliation{$^1$ Center for Theoretical Physics, Zewail City for Science and Technology, 6 October City, Cairo, Egypt.\\
$^2$Department of Mathematics, Faculty of Science,  Ain Shams University, Cairo, Egypt.\\
$^3$School of Physics and Astronomy, University of Southampton,
Highfield, Southampton SO17 1BJ, UK.\\
$^3$Department of Physics, Faculty of Science, Al Azhar
University, Cairo, Egypt. }
\date{\today}
\abstract{We show that the low energy effective model derived from
$SU(5)$ with $45$-plet Higgs can account for the recently reported
enhanced diphoton decay rate of the Standard Model (SM)-like Higgs
with mass about 125 GeV. This model extends the SM by an extra
Higgs doublet and color-octet scalar doublet. We show that the
charged octet scalars are not severely constrained by flavor
changing neutral current and can be light. However, the
$K^0-\bar{K}^0$ mixing implies that the neutral octet-scalar mass
should be larger than $400$ GeV for $\tan \beta \sim {\cal O}(1)$.
The role of charged octet scalar in the loop of Higgs decay into
diphoton is investigated . We point out that the most significant
impact of this model on the diphoton decay width comes from the
suppression of top-quark coupling with SM-like Higgs or even
flipping its sign that leads to important enhancement in $\Gamma(h
\to \gamma \gamma)$. We also study the implications of the neutral
octet-scalar contributions to the gluon fusion Higgs production
cross section in alleviating the apparent tension between
enhancement of diphoton decay rate and suppression of $\sigma(gg
\to h)$.}

\maketitle
%----------------------------------------------------------
\section{Introduction}

The Grand Unified Theory (GUT) $SU(5)$, which was proposed by
Georgi and Glashow in 1974\cite{Georgi:1974sy}, is the simplest
extension of the Standard Model (SM) that provides a natural
framework for the unification of fundamental interactions. In this
model, the SM gauge groups $SU(3)_C \times SU(2)_L \times U(1)_Y$
are unified into a single simple gauge group and the SM quarks and
leptons are combined into irreducible $SU(5)$ representations:
$\bar{5}$ and $10$, namely%
\bea%
5^* &=& (\bar{3},1)_{1/3}  \oplus (1,2)_{-1/2}  \equiv (d^c, \tilde{l}),\\
10 &=& (3,2)_{1/6} \oplus (\bar{3},1)_{-2/3} \oplus (1,1)_{1}
\equiv (q, u^c, e^c),%
\eea %
where $\tilde{l}= i \tau_2 l$ with $l=(\nu,e)^T$ and $q=(u,d)^T$.
The Higgs sector in the minimal $SU(5)$ contains adjoint
24-dimensional scalar field representation to break $SU(5) \to
SU(3)_C \times SU(2)_L \times U(1)_Y$ and fundamental
5-dimensional scalar field representation to break $SU(2)_L \times
U(1)_Y \to U(1)_{em}$.

The minimal $SU(5)$ is very predictive: It naturally predicts the
quantization of electric charge, where ${\rm Tr}Q_{\bar{5}}=0$,
hence $Q(d) = \frac{1}{3} Q(e^-)$. It also predicts
$\sin^2\theta_W$ in a very good agreement with the current result
\cite{Buras:1977yy}. It leads to a bottom-tau mass ratio at low
energy: $m_b/m_\tau \simeq 3$, which is consistent with the
measured masses. However the minimal $SU(5)$ has several
drawbacks: It predicts proton decay, $p \to e^+ \pi^0$, with a
life-time of order $10^{32}$ years, which is in a clear
contradiction with the experimental bound $\gsim 5 \times 10^{33}$
\cite{Nishino:2009aa}. It implies a wrong mass relation:
$m_e/m_\mu = m_d/m_s$. Furthermore, in minimal $SU(5)$ the gauge
couplings do not unify and the model suffers from a naturalness
problem due to the gauge hierarchy problem and a doublet-triplet
splitting. Finally, it does not predict right-handed neutrinos,
therefore the neutrinos are massless in minimal $SU(5)$ which
contradicts the recent neutrinos oscillation results.

There have been several efforts to construct more realistic
extensions of the minimal $SU(5)$ through imposing an extra
symmetry or extending the fermion and/or Higgs sector
\cite{Georgi:1979df,Nandi:1980sd,Kalyniak:1982pt,Bajc:2002bv,Aristizabal:2003zn,Dorsner:2005ii,Perez:2007iw}.
For instance, it has been shown that the fermion mass relation,
proton stability, and gauge unification can be adjusted by
extending the $SU(5)$ Higgs section and introducing another scalar
field in the $45$-representation
\cite{Georgi:1979df,Nandi:1980sd,Kalyniak:1982pt}. One of the
salient features of $SU(5)$ model with $45$-plet scalar is that
the resultant low energy effective model is the SM-like with two
Higgs doublets and light octet scalar. This rich Higgs sector has
the potential to account for the ATLAS and CMS recent experimental
result for $H \to \gamma \gamma$ signal strength, which is $\sim
1.5$ times larger the SM prediction
\cite{Aad:2012tfa,Chatrchyan:2012ufa}. This excess is very
interesting and it provides a good hint of a possible new physics.

The latest results of ATLAS and CMS collaborations, announced in
Morioned conference \cite{moriond}, confirmed the Higgs discovery
with mass of order 125 GeV. Both collaborations have independently
performed search for the Higgs boson in different decay channels.
The most confirmed discovery channels are $H \to \gamma \gamma$,
$H \to ZZ^{(*)} \to 4 l$, and $H \to WW^{(*)} \to l \nu l \nu$ at
integrated luminosities of $5.1~ fb^{-1}$ taken with energy
$\sqrt{s} =7$ TeV and $19.6~ fb^{-1}$ taken at $\sqrt{s}=8$ TeV.
While ATLAS confirmed the excess in $H \to \gamma \gamma$ and
found that the best fit of signal strength is given by%
\be%
\sigma/\sigma_{SM} = 1.65 \pm 0.24~ {\rm (stat)} ^{+0.25}_{-0.18}
{\rm (syst)}, %
\ee%
the CMS changed their previous results from
$\sigma/\sigma_{SM}=1.56 \pm 0.43$ to $1.11\pm 0.31$ with cut
based events and $0.78 \pm 0.27$ with selected and categorized
events. Also ATLAS experiment has reported an excess in $ H \to
ZZ^{(*)} \to 4 l$ as well with $\sigma/\sigma_{SM}= 1.5 \pm 0.4 $
for $m_H =125.5$ GeV. On the contrary CMS experiment showed that
this channel is consistent with the SM expectation with
$\sigma/\sigma_{SM}= 0.91^{+0.30}_{-0.24}$. Finally for $H \to
WW^{(*)} \to l \nu l \nu$, ATLAS showed that the signal strength
of this process at $m_H=125$ is $1.01 \pm 0.21 (stat)\pm 0.12
(syst)$ and CMS found that it is given $0.76 \pm 0.13 (stat) \pm
0.16 (syst)$. From these results, it is clear that much further
analysis is needed to reveal the discrepancy between the results
of the two experiments. Also it seems that these results are not
entirely consistent with the SM predictions, particularly
$R_{\gamma \gamma}$. In our effective $SU(5)$ model, we may have
significant contributions to $H \to \gamma \gamma $ decay from the
loop of charged octet scalar $(S^\pm)$. In addition, the neural
octet scalar $(S^0)$ may have important effect on the Higgs
production through the gluon fusion $gg \to H$. Therefore, the
expected tension between the enhancement of $\Gamma(H \to \gamma
\gamma)$ and the associated suppression of $\sigma(gg \to H)$
\cite{Buckley:2012em} can be relaxed.

The possibility of light octet scalars arises in several
extensions of the SM \cite{light-octet,Burgess:2009wm}. The
potential discovery of these particles at the LHC has been
investigated in Ref.\cite{LHC-octet}. It is interesting to note
that the most robust constraints on the octet scalar masses are
due to the direct searches for pair of octet scalars at the LHC,
which lead to \cite{ATLAS:2012ds,Aaltonen:2013hya}: $M_{S}
> 287$ GeV at $95\%$ confidence level. In
most of octet scalar analysis, the Minimal Flavor Violation (MFV)
was assumed. In this case, the Yukawa couplings of octet scalar to
the SM quarks are proportional to the Yukawa couplings of the
SM-like Higgs to quarks, {\it i.e.}, $Y_{Sqq'} = {\rm cosnt.}
\times Y_{Hqq'}$. In general, this assumption is not true and
$S^{\pm}$ and $S^0$ can be source of new flavor violations beyond
the SM ones, which are proportional to the quark mixing $V_{CKM}$.

In this paper, we derive the constraints on the octet scalar
masses due to the experimental bounds of $K^0 -\bar{K}^0$ mixing
and $B^0_q - \bar{B}^0_q$ mixing, where $q= d, s$. We show that
the constraints imposed on neutral octet scalars from the current
flavor violation limits may be more stringent than the direct
search constraints. While the charged octet scalars are
essentially free of flavor changing neutral current constraints,
they should be just heavier than $W^\pm$ gauge boson to give
contribution less than the SM one. In this respect, we show that
$H \to \gamma \gamma$ decay can be enhanced through the
contribution of the charged octet scalars, so that the above
mentioned experimental results can be accommodated.

This paper is organized as follows. In section 2, we  briefly
review $SU(5)$ with $45$-plet. In particular, we analyze the Higgs
sector at low energy and emphasize that it consists of two Higgs
doublets with neutral and charged octet scalars. The interactions
of the octet scalars with the SM particles are also provided. In
section 3, we study the constraints imposed on the octet scalar
masses from $K^0 - \bar{K}^0$ and $B^0_q -\bar{B}^0_q$ mixing,
where $q=d,s$, in addition to the constraints obtained from the
direct search at Tevatron and LHC. Section 4 is devoted for the
octet scalar contribution to the decay rate of $H \to \gamma
\gamma$ and enhancing the signal strength $R_{\gamma \gamma}$.
Finally, our conclusions are given in section 5.

%----------------------------------------------------------

\section{$SU(5)$ with $45$-plet}

The Higgs fields in this class of $SU(5)$ is composed of $24_H,
5_H$ and $45_H$ representations, where $24_H$ Higgs fields acquire
vacuum expectation value (vev) at GUT scale and break $SU(5)$
group down to the SM. The $5_H$ and $45_H$ Higgs fields contribute
in the electroweak symmetry breaking of the SM. In this case, the
$SU(5)$ invariant
Yukawa Lagrangian is given by%
\bea%
{\cal L}_{\rm Yuk}= Y_1 \bar{5}_{\alpha} 10^{\alpha\beta}
(5^*_{H})_\beta + Y_2 \bar{5}_{\delta} 10^{\alpha\beta}
(45^{*}_H)_{\alpha \beta}^\delta +
\epsilon_{\alpha\beta\gamma\delta \lambda} \Big[Y_3
10^{\alpha\beta} 10^{\gamma\delta} 5_H^\lambda + Y_4
10^{\alpha\beta} 10^{\xi \gamma}_L (45_H)^{\delta \lambda}_\xi
\Big].~%
\label{yuk}
\ee %
It is worth remembering that the Higgs bosons $5_H$ and
$45_H$
transform under the SM gauge as%
\bea %
5_H &=& (3,1)_{-1/3} \oplus (1,2)_{1/2} \\
45_H &=& (8,2)_{1/2}\oplus (1,2)_{1/2}\oplus (3,1)_{-1/3}\oplus
(3,3)_{-1/3}\oplus (6^*,1){-1/3}\oplus (3^*,2)_{-7/6}\oplus
(3^*,1)_{4/3}.%
\eea%
Also $45_H$ satisfies the following constraints
\cite{light-octet}: $45^{\alpha \beta}_\gamma = - 45^{\beta
\alpha}_\gamma$ and $\sum_\alpha^5 (45)^{\alpha \beta}_\alpha =0$.
Thus, the electroweak symmetry $SU(2)_L \times U(1)_Y$ can be
spontaneously broken into $U(1)_{em}$ through the non-vanishing
vev of the doublets in $5_H$ and
$(45_H)^{\alpha 5}_\alpha$, namely%
\bea%
\langle 5_H \rangle &=& v_5,\\
\langle 45_H \rangle^{15}_1 &=& \langle 45_H \rangle^{25}_2 =
\langle 45_H \rangle^{35}_3 = v_{45}, \hspace{0.5cm} \langle 45_H
\rangle^{45}_4 =
-3 v_{45}.%
\eea

The $5_H$-doublet is defined as%
\be%
H \equiv (1,2)_{1/2} = \left(
                       \begin{array}{c}
                         H^+ \\
                         H^0 \\
                       \end{array}
         \right), \ee
while the $45_H$-doublet can be written as%
\bea \nonumber && D \equiv (1,2)_{1/2}  = \left(
                           \begin{array}{cc}
                             D^{54}_4  &  D^{54}_5 \\
                              D^{45}_4 & D^{45}_5 \\
                           \end{array}
                         \right) = \left(
             \begin{array}{c}
               - D^{45}_5 \\
                 D^{45}_4 \\
             \end{array}
           \right) =  \left(
                        \begin{array}{c}
                          - D^+ \\
                          D^0 \\
                        \end{array}
                      \right) \eea

In addition, the $45_H$-color octet scalars are defined as \cite{light-octet}%
\be%
S^{ia}_j\equiv (8,2)_{1/2} = (45_H)^{ia}_{j} -
\frac{1}{3}\delta^i_j (45_H)^{ma}_{m} = \left(
                                 \begin{array}{c}
                                   S^{+} \\
                                   S^{0}_R+iS^{0}_I \\
                                 \end{array}
                               \right) \equiv S^A T^A, %
\ee%
where $i,j= 1,2,3$, $A= 1, . . , 8$, and $T^A$ are the $SU(3)$
generators. It clear that the octet scalars are defined such that
they have vanishing vevs. In this case, one can easily show that the fermions masses are given by \cite{Georgi:1979df,Nandi:1980sd} %
\bea%
M_E &=& Y_1^T  v^*_5 - 6 Y_2^T v^*_{45},\label{mE}\\
M_D &=& Y_1 v^*_5 + 2 Y_2 v^*_{45}, %
\label{mD}
\\
M_U &=& 4 (Y_3 + Y_3^T) v_5 - 8 (Y_4^T - Y_4) v_{45}.%
\label{mU}%
\eea %
Thus, the usual $SU(5)$ wrong mass relation between the masses of
charged leptons and down quarks is resolved. In addition, from
Eq.(\ref{mU}) one notices that the up-quark masses may depend only
on $v_5$ and the Yukawa couplings $Y_3$ if the Yukawa matrix $Y_4$
is symmetric or if one adopted the common bases where the up quark
mass matrix is diagonal and down quarks are the only source for
the quark mixing matrix $V_{CKM}$.

Also, it is worth noting that both $5_H$ and $45_H$ Higgs fields
are coupled with $24_H$ Higgs fields that acquire vevs of order
GUT scale. Therefore, it is a challenging to keep the Higgs
doublets of $5_H$ and $45_H$, in addition to the octet scalar of
$45_H$ light (of order the electroweak scale) while all other
fields are quite heavy. This is known as splitting problem. In
minimal $SU(5)$, this problem was to justify the splitting between
the doublet and triplet of $5_H$. Now we have an additional
splitting among the doublet, octet of $45_H$ and its triplets,
sixet. In principle, the potential $V(24_H, 5_H)$ and $V(24_H,
45_H)$ contain several free parameters that can be adjusted such
that the required pattern is obtained. In this case the low energy
effective model derived from the non-minimal $SU(5)$ is the SM
extended by an extra Higgs doublet and neutral and charged octet
scalars.

\subsection{$SU(5)$-two Higgs doublets}
The $SU(5)$ invariant potential of $5_H$ and $45_H$ Higgs fields
is given by%
\bea%
\nonumber V(5_H, 45_H) =&& -\mu^2_5\; 5^*_\alpha 5^\alpha +
\lambda_1\; (5^*_\alpha 5^\alpha)^2 -\mu^2_{45}\;
45^{\gamma*}_{\alpha\beta} 45^{\alpha\beta}_\gamma + \lambda_2 \;
(45^{\gamma*}_{\alpha\beta} 45^{\alpha\beta}_\gamma)^2 +
\lambda_3\; (45^{\gamma*}_{\alpha\beta} 45^{\alpha\beta}_\gamma)\;
5^*_\delta 5^\delta    \\&& + \lambda_4\;
45^{\gamma*}_{\alpha\beta}\ 5^\beta\ 5^*_\delta\
45^{\alpha\delta}_\gamma +\frac{1}{2} \lambda_5\;  \Big[
5^*_\beta\ 45^{\alpha\beta}_\gamma\ 5^*_\delta\
45^{\gamma\delta}_\alpha + 45^{\gamma*}_{\alpha\beta}\ 5^\beta\
45^\alpha_{\gamma\delta}\ 5^\delta \Big] + \lambda_6\;
45^{\alpha\beta}_\gamma 5^\gamma 5^*_\delta
45_{\alpha\beta}^{*\delta}  .\nonumber %
\eea
After $SU(5)$ symmetry breaking, one finds%
\bea%
V(H,D) &&= -\mu_H^2 H^\dagger H + \lambda_1 (H^\dagger H)^2 -
\mu_D^2 D^\dagger D + \lambda_2 (D^\dagger D)^2 + \lambda'_3
(D^\dagger D) (H^\dagger H) +\frac{1}{2}\lambda_5 [( H^\dagger
D)^2 + (D^\dagger H )^2 ]\nonumber\\&&  + \lambda'_6\
(\widetilde{D} H)\  (\widetilde{D} H)^\dagger, %
\eea %
where $\lambda'_3= 2\lambda_3+\lambda_4,$\ $\lambda'_6=2\lambda_6$
and $\widetilde{D}=i\tau_2 D$. Therefore, the scalar potential of
neutral Higgs bosons is given by%
\bea %
\nonumber V(H^0,D^0) &&= -\mu_H^2 H^{0*} H^0 + \lambda_1
(H^{0*} H^0)^2 - \mu_D^2 D^{0*} D^0 + \lambda_2 (D^{0*} D^0)^2 +
\lambda'_3 (D^{0*} D^0) (H^{0*} H^0)\\&& +\frac{1}{2}\lambda_5 [(
H^{0*} D^0)^2 + (D^{0*} H^0 )^2 ].%
\eea%
These neutral components develop vacuum expectations values:
$\langle H^0 \rangle = v_1 \equiv v_5$ and $\langle D^0 \rangle =
v_{2} \equiv -3 v_{45}$. As in two Higgs doublet models, the mass
of the $W$-gauge bosons is given by $M_W = g v$, where $v=
\sqrt{v_1^2 + v_{2}^2}$ and one defines $\tan \beta= v_{2}/v_1$.
In addition, the following minimization conditions
are obtained:%
\bea%
&& -\mu_H^2   + 2 \lambda_1\ v^2_1   + (\lambda'_3+ \lambda_5)
v^2_{2} =0, \\
&& -\mu_D^2   + 2 \lambda_2\ v^2_{2} + (\lambda'_3+ \lambda_5)
v^2_{1} =0. %
\eea%

In order to obtain the physical Higgs fields and their masses, one
should write the two doublets $H$ and $D$ around their vacua as
follows: %
\bea%
H &=& (H^+, H^0) = (H^+, v_5 + H^0_R + i H^0_I) , \\
D &=& (-D^+, D^0) = (-D^+, v_{45} + D^0_R + i D^0_I) ,%
\eea%
where the real components correspond to the CP–even Higgs bosons
and the imaginary components correspond to the CP–odd Higgs and
the Goldstone boson. The mass matrix of the CP-even Higgs can be
obtained as%
\be%
M_R^2= \left(\begin{array}{cc}
    -\mu^2_H+6\lambda_1 v_1^2+ \lambda v^2_{2} & \lambda v_1 v_{2} \\
      \lambda v_1 v_{2}  & -\mu^2_D+4\lambda_2 v_{2}^2+\lambda v^2_{1}
\end{array}
\right) = \left(\begin{array}{cc}
   4\lambda_1 v_1^2 ~~ &  ~~ \lambda v_1 v_{2}\\
   \lambda v_1 v_{2} ~~ & ~~  4\lambda_2 v_{2}^2
\end{array}
\right),   %
\ee%
with $\lambda = \lambda'_3+\lambda_5$. The last equality is
obtained by using the minimization conditions to write $\mu^2_H$
and $\mu^2_D$ in terms of $v_5$ and $v_{45}$. Therefore, the mass
eigenstates fields $h$ and $H$ are given as%
\be%
\left(\begin{array}{c}
H\\
h \end{array}\right)= \left(\begin{array}{cc} \cos \alpha ~~ & ~~
\sin \alpha\\
- \sin \alpha ~~ & ~~ \cos\alpha \end{array}\right)
\left(\begin{array}{c}
H^0_R\\
D^0_R \end{array}\right),%
\ee%
where the mixing angle $\alpha$ is defined by%
\be%
\tan 2 \alpha = \frac{\lambda v_1 v_{2}}{2(\lambda_1 v_1^2-
\lambda_2 v_{2}^2)}.%
\ee%
The masses of the CP–even Higgs bosons $h$ and $H$ are given by
\be%
M^2_{h, H}= 2\lambda_1 v_1^2+ 2 \lambda_2 v_{2}^2 \mp
\sqrt{(2\lambda_1
v_1^2 -2\lambda_2 v_{2}^2 )^2  + \lambda^2 v_1^2 v_{2}^2}.%
\ee%
It is clear that $\lambda$ is the mixing parameter between the
SM-like $h$ Higgs and the extra Higgs $H$. For $\lambda= 0$, there
is no mixing and the Higgs masses are given by $m_h = 2 \sqrt{
\lambda_1} v_1$ and $m_H = 2\sqrt{\lambda_2} v_{2}$.

Similarly the mass matrix of the CP-odd Higgs and Goldstone boson
can
be obtained as%
\be%
M_I^2= \left(\begin{array}{cc}
    -\mu^2_H + 2\lambda_1 v_1^2+\lambda v^2_{2} & 2\lambda_5 v_1 v_{2}  \\
      2\lambda_5 v_1 v_{2}  & -\mu^2_D+ 2\lambda_2 v_{2}^2+\lambda v^2_{1}
\end{array}
\right) = \left(\begin{array}{cc}
   -2\lambda_5 v_{2}^2 ~~ &  ~~ 2 \lambda_5 v_1 v_{2}\\
   2 \lambda_5 v_1 v_{2} ~~ & ~~  -2\lambda_5 v_{1}^2
\end{array}
\right).   %
\ee%
Since the determinant of $M_I^2$ is zero, one eigenvalue vanishes
and corresponds to the Goldstone boson mass, while the other
eigenvalue corresponds to the pseudoscalar Higgs, $A = - \sin
\beta H_0^I + \cos \beta D^0_I$, with mass given
by %
\be%
M^2_{A} = 2 \lambda_5(v_1^2 + v_{2}^2)= 2 \lambda_5 v^2. %
\ee %

Finally, the mass matrix of the charged Higgs bosons is given by
\bea%
M_{H^\pm}^2 &=& \left(\begin{array}{cc}
    -\mu^2_H + 2\lambda_1 v_1^2 + (\lambda'_3+\lambda'_6)  v^2_{2} & (-\lambda_5+\lambda'_6) v_1 v_{2} \\
      (-\lambda_5+\lambda'_6) v_1 v_{2}  & -\mu^2_D+2\lambda_2 v_{2}^2+ (\lambda'_3+\lambda'_6) v^2_{1}
\end{array}
\right)\nonumber\\
&=& \left(\begin{array}{cc}
   (-\lambda_5+\lambda'_6) v_{2}^2 ~~ &  ~~ (-\lambda_5+\lambda'_6) v_1 v_{2}\\
   (-\lambda_5+\lambda'_6) v_1 v_{2} ~~ & ~~  (-\lambda_5+\lambda'_6)  v_{1}^2
\end{array}
\right).   %
\eea%
One of the eigenvalues of this mass matrix is zero and corresponds
to a massless charged Goldstone, while the other eigenvalue
corresponds a charged Higgs boson, $H^\pm = -\sin \beta H^\pm +
\cos \beta D^\pm$, with mass given by
by%
\be %
M^2_{H^\pm} = (\lambda_6 - \lambda_5) v^2.%
\ee%

Now we consider the induced Yukawa couplings of the lightest
neutral Higgs scalar ``SM-like Higgs" to the SM fermions. From the
$SU(5)$ Yukawa interactions in Eq.(\ref{yuk}), the Higgs doublets
$H$ and $D$ have the following low energy scale interactions with
the SM fermions.%

\be {\cal L}=  Y_1^T \bar{e}_R H^{\dagger} l_L +\ 2 Y^T_2
\bar{e}_R D^\dagger  l_L +\  Y_1 \bar{d}_R H^{\dagger} Q_L + Y'_3
\epsilon_{\alpha \beta}  \bar{u}_R Q^\alpha_L  H^\beta + Y'_4
\epsilon_{\alpha \beta}  \bar{u}_R  Q^\alpha_L D^\beta + h.c.,
\ee%
where $Y'_3 = 4 (Y_3 + Y_3^T)$ and $Y'_4 = 4 Y_4$. This indicates
that although $Y_4$ may not contribute in the up quark mass matrix
in certain cases, it has an important effect on their interactions
with Higgs particles. Also one can express $Y_1$ and $Y_2$ Yukawa
couplings from the fermion mass relations in terms of the down
quark and charged lepton
masses:%
\be%
Y_1 = \frac{3M_D+M_E}{4v_5}, \hspace{0.75cm}
Y_2=\frac{M_D-M_E}{8v_{45}}.%
\ee%
Therefore, the Yukawa Lagrangian can be expressed in the physical
basis as %
\bea \nonumber {\cal L}=&& \bar{e}_R \Big[-\Big(\frac{3m_D
V^\dagger_{CKM}+m_E}{4v_5}\Big) \sin \beta  + \Big(\frac{m_D
V^\dagger_{CKM} -m_E}{4v_{45}}\Big) \cos \beta \Big] H^- \nu_{eL}
\nonumber\\&& +\bar{e}_R\Big[\Big( \frac{3m_D
V^\dagger_{CKM}+m_E}{4v_5}\Big) (H \cos \alpha - h \sin \alpha -i
A \sin \beta) \nonumber\\
&&~~~~~~~+ \Big(\frac{m_D V^\dagger_{CKM}-m_E}{4 v_{45}}\Big)(H
\sin \alpha + h \cos \alpha + i A \cos \beta ) \Big] e_L
\nonumber\\&&+ \bar{d}_R \Big[-\Big( \frac{3m_D
V^\dagger_{CKM}+m_E}{4 v_5}\Big) \sin \beta \Big]H^- u_L
\nonumber\\
&&+ \bar{d}_R \Big[\Big( \frac{3m_D + m_E V_{CKM}}{4 v_5}\Big) (H
\cos \alpha - h \sin \alpha -i A \sin \beta) \Big]\;d_L
\nonumber\\&&+ \bar{u}_R \Big[- Y'_3 \sin \beta + Y'_4 \cos \beta
\Big] V_{CKM} H^+ d_L \nonumber\\&& +\bar{u}_R \Big[Y'_3  (H \cos
\alpha  - h \sin  \alpha -iA \sin \beta) + Y'_4 (H \sin \alpha + h
\cos \alpha +iA \cos \beta ) \Big] u_L +h.c., \eea
If one assumes flavor diagonal charged leptons and up-quarks,
while the down quark mass matrix is diagonalized by left-handed
rotation only, {\it i.e.}, $V_L^d = V_{CKM}$ and $V_R^d=I$. In
this case, one can summarize the SM-like Higgs
couplings to the SM fermions as follows:%
\bea %
&&Y_{huu}= - \ Y'_3 \sin \alpha + Y'_4 \cos \alpha= - \frac{m_U}{v} \frac{\sin\alpha}{\cos\beta} + Y'_4 \cos\alpha,\\
&&Y_{hdd}= - \Big( \frac{3m_D + m_E.V_{CKM}}{4v_5}\Big) \sin\alpha,\\
&& Y_{hee}= - \Big( \frac{3m_D V^\dagger_{CKM}+m_E}{4v_5}\Big)
\sin \alpha + \Big(\frac{m_D
V^\dagger_{CKM}-m_E}{4v_{45}}\Big)\cos
\alpha . %
\eea

Similarly, one can derive the Higgs couplings to the electroweak
gauge bosons. These coupling are obtained from the kinetic terms
of the fields $H$ and $D$ in the Lagrangian%
\be {\cal L}_{kin} = (D^\mu H)^\dagger (D_\mu H) + (D^\mu
D)^\dagger
(D_\mu D).%
\ee%
Expanding the covariant derivative $D_\mu$ and performing the
usual transformations on the gauge and scalar fields to obtain the
physical fields, one can identify the couplings between the Higgs
and gauge bosons. In particular, the coupling of the SM-like Higgs
to $
W_\mu^+ W^-_\nu$  and to $ Z_\mu Z_\nu$ are given by%
\bea%
g_{hW^+ W^-} &\equiv & g M_W \sin(\beta - \alpha), \\
g_{h Z Z} &\equiv & \frac{g M_z }{\cos \theta_W}  \sin(\beta -  \alpha) . \eea

\begin{figure}[t]
\begin{center}
\epsfig{file=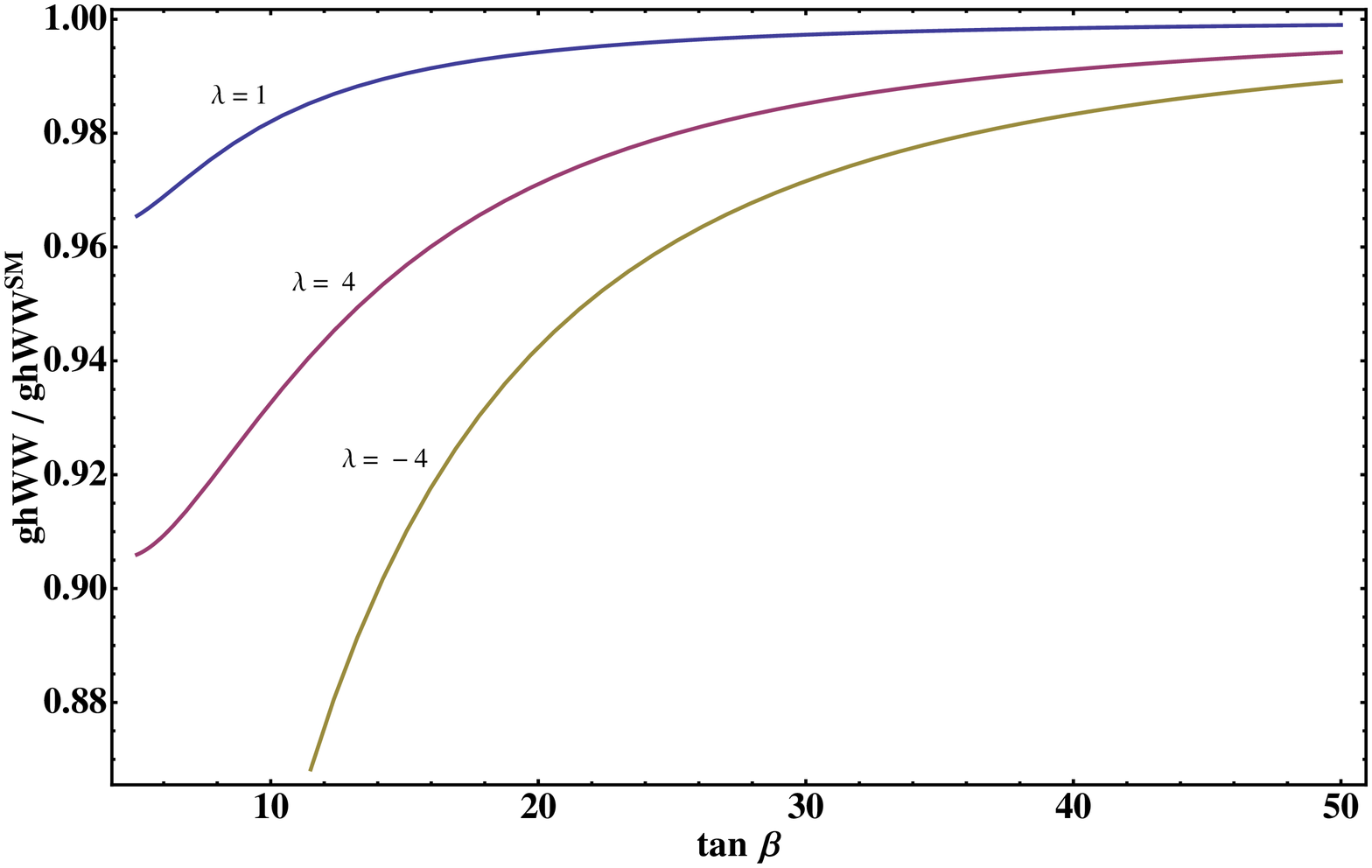,height=5.5cm,width=7.0cm,angle=0}~~~~~~
\epsfig{file=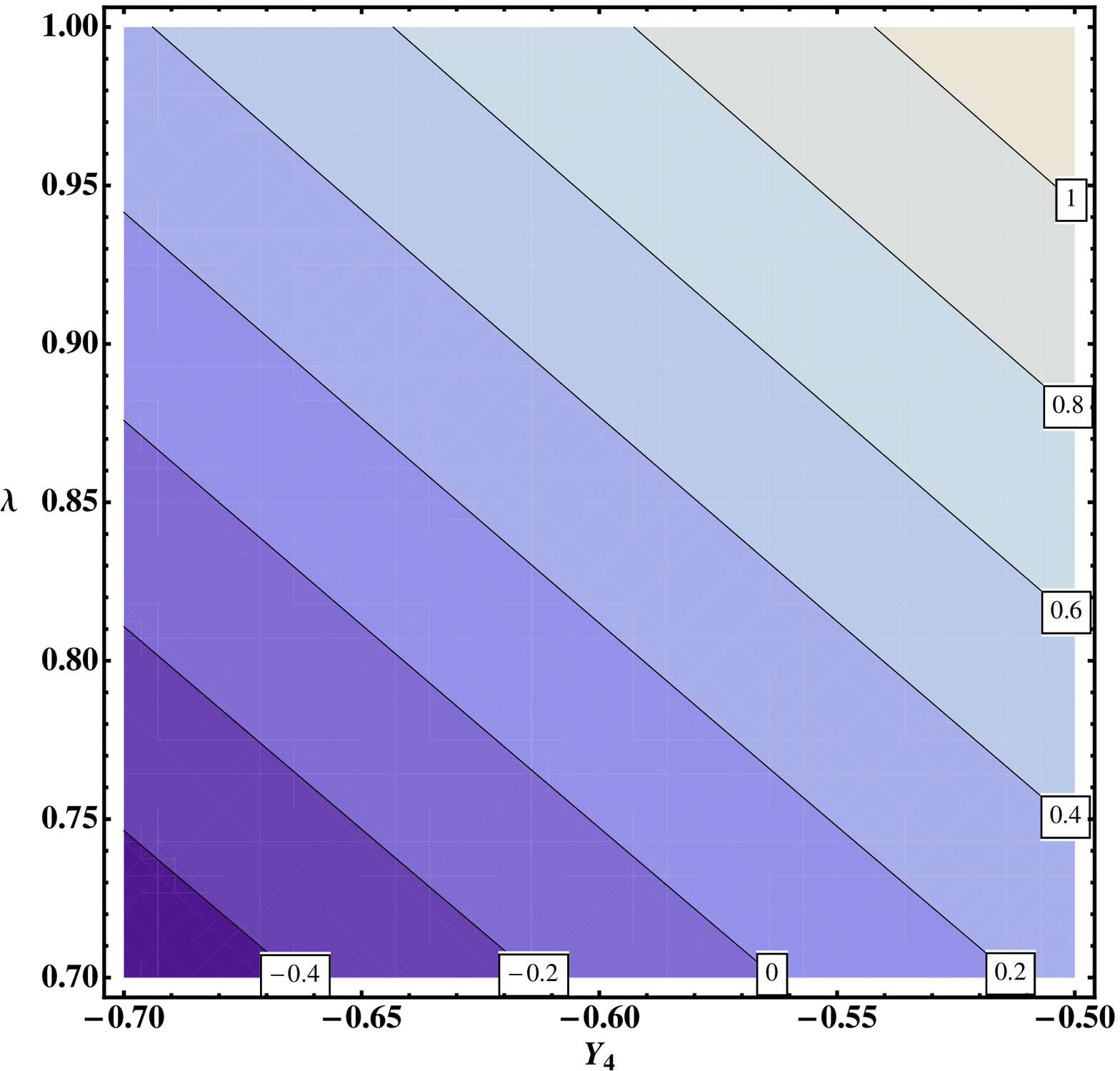,height=5.5cm,width=7.5cm,angle=0}
\caption{\label{ghww} (Left panel): The ratio
$g_{hWW}/g_{hWW}^{SM}$ as a function of $\tan \beta$ for $\lambda=
1, \pm 4$, where $\lambda= 2 \lambda_3 +\lambda_4 + \lambda_5$ as
defined in Eqs.(2.11)-(2.17). (Right panel): Contour plot of the
ratio $Y_{htt}/Y_{htt}^{SM}$ as a function of $\lambda$ and the
coupling $Y_4$ of $45$-doublet with the top-quark. }
\end{center}
\end{figure}

In Fig. \ref{ghww} we display the ratio of $g_{hW^+W^-}$
normalized by the SM coupling $g_{hW^+W^-}^{SM} = g M_W$ as
function of $\tan \beta$. As can be seen from this plot, large
values of $\lambda$ with small $\tan \beta$ lead to $g_{hW^+W^-}$
smaller than the SM result. Note that $\lambda$ is defined as
$\lambda= 2 \lambda_3 +\lambda_4 + \lambda_5$, hence it can vary
from -4 to +4. However, small $g_{hW^+W^-}$ is not favored, if we
are interested in enhancing $\Gamma(h\to \gamma \gamma)$ respect
to the SM expectation \cite{Aad:2012tfa,Chatrchyan:2012ufa}.
Therefore, one may consider the following constraints: $\lambda
\sim {\cal O}(1)$ or $\tan \beta$ is quite large. In this figure
we also provide a contour plot for $Y_{htt}/Y_{htt}^{SM}$ as a
function of $\lambda$ and the coupling $Y_4$. As can be seen from
this figure for a large region of parameter space $Y_{htt} <
Y_{htt}^{SM}$ is obtained. It is also remarkable that $Y_{htt}$
may flip its sign and becomes negative. In this case, as we will
discuss in the next section, the top contribution to $h \to \gamma
\gamma$ will have a constructive interference with
$W$-contribution that leads to enhancement of $\Gamma(h \to \gamma
\gamma)$.

\subsection{Octet scalar Interactions}
The interaction of octet scalars with the gluon is one of their
most relevant interactions with the SM particles. This interaction
is obtained from the kinetic term of $45_H$: ${\rm Tr}
\left[(D_\mu 45_H)^\dag (D^\mu 45_H) \right]$, where the covariant
derivative of $(45_H)^{\alpha \beta}_{\gamma}$ is
give by %
\be%
D_\mu (45_H)^{\alpha \beta}_{\gamma} = \partial_\mu (45_H)^{\alpha
\beta}_{\gamma} - i g (A_\mu)^\alpha_\lambda (45_H)^{\lambda
\beta}_{\gamma} - i g (A_\mu)^\beta_\lambda (45_H)^{\alpha
\lambda}_{\gamma} - i g (A_\mu)^\lambda_\gamma (45_H)^{\alpha
\beta}_{\lambda}, %
\ee%
where $A_\mu \equiv A_{\mu}^A T^A$ is the $SU(5)$ gauge bosons.
This leads to the
following covariant derivative of octet scalars $S^{ia}_j$:%
\be %
{\cal D}_\mu S^{ia}_j = \partial_\mu S^{ia}_j - i \frac{g_s}{2}
(G^\alpha_\mu \lambda^\alpha)^{ik}\  S^{ka}_j - i\frac{g_s}{2}
(G^\alpha_\mu \lambda^\alpha)^{k}_j\ S^{ia}_k  - i \frac{g}{2}
(A_\mu^r \tau^r)^{ab}\ S^{ib}_j -i \frac{g'}{2} Y B_\mu  S^{ia}_j.
\ee %
Therefore, one can extract the follow interactions between gluons
and scalar octets:
\bea%
{\cal L}^S_{gluon}&=&i \frac{g_s}{2} \Big[ S^{j-}_k (G^\mu
\lambda)^*_{ki} \partial_\mu S^{i+}_j + S^{j0}_k (G^\mu
\lambda)^*_{ki} \partial_\mu S^{i0}_j +  S^{k-}_{i} (G^\mu
\lambda)^{j*}_k \partial_\mu S^{i+}_j +S^{k0}_{i} (G^\mu T)^{j*}_k
\partial_\mu S^{i0}_j \Big] \nonumber\\
&+& \frac{g^2_s}{4} \Big[S^{j-}_k (G^\mu \lambda)^*_{ki} (G_\mu
\lambda)^{il} S^{l+}_j + S^{j0}_k (G^\mu \lambda)^*_{ki} (G_\mu
\lambda)^{il} S^{l0}_j + S^{k-}_{i} (G^\mu \lambda)^{j*}_k (G_\mu
\lambda)^s_j S^{i+}_s \nonumber\\
&+&S^{k0}_{i} (G^\mu \lambda)^{j*}_k (G_\mu \lambda)^s_j S^{i0}_s
+ S^{k-}_{i} (G^\mu \lambda)^{j*}_k (G_\mu \lambda)^{il} S^{l+}_j
+ S^{k0}_{i} (G^\mu \lambda)^{j*}_k (G_\mu \lambda)^{il} S^{l0}_j
\Big], \eea
which can be written as %
\bea %
\nonumber  {\cal L}^S_{gluon} &&= i g_s {\rm Tr}\Big[ S^{A-} T^A
G^{\mu B} T^B
\partial_\mu  S^{D+} T^D + S^{A0}_R T^A G^{\mu B}  T^B
\partial_\mu  S^{D0}_R T^D + S^{A0}_I T^A G^{\mu B}  T^B
\partial_\mu  S^{D0}_I T^D  \Big] \nonumber\\
&+& g_s^2\ {\rm Tr}\Big[ S^{A-} T^A  G^{\mu B}  T^B G^{C}_\mu
T^{C}  S^{D+} T^D + S_R^{A0} T^A  G^{\mu B} T^B
G^{C}_\mu T^{C}  S^{D0}_R T^D \nonumber\\
&+& S^{A0}_I T^A  G^{\mu B} T^B G^{C}_\mu T^{C} S^{D0}_I T^D \Big]
+ h.c. \eea
\bea \nonumber  {\cal L}^S_{gluon} &&= i g_s {\rm Tr}\Big[ S^{A-}
G^{\mu B} \partial_\mu  S^{D+} + S^{A0}_R G^{\mu B}
\partial_\mu  S^{D0}_R + S^{A0}_I\ G^{\mu B}  \partial_\mu  S^{D0}_I \Big]{\cal F}^{ABD} \nonumber\\
&+& g_s^2 {\rm Tr}\Big[ S^{A-}  G^{\mu B}  G^{C}_\mu S^{D+} +
S_R^{A0} G^{\mu B}  G^{C}_\mu  S^{D0}_R  + S^{A0}_I  G^{\mu B}
G^{C}_\mu  S^{D0}_I \Big]{\cal F}^{ABCD}+ h.c . ~~~%
\eea%
With %
\bea%
{\cal F}^{ABD}&=& tr[T^A T^B T^D] = 1/4 \Big( d^{ABD} + i f^{ABD}
\Big) ,\\%
{\cal F}^{ABDE} &=& tr[T^A T^B T^D t^E] =\frac{2}{9} \delta^{AB}
\delta^{DE} +\frac{1}{8} \Big[ d^{ABC} d^{DEC} +i d^{ABC} f^{DEC}
+i f^{ABC} d^{DEC} - f^{ABC} f^{DEC} \Big],\nonumber\\%
\eea %
where $d^{ABC}$ and $f^{ABC}$ are the $SU(3)$ symmetric and
antisymmetric structure constants, respectively.

\begin{figure}[t]
\begin{center}
\includegraphics[scale=0.8]{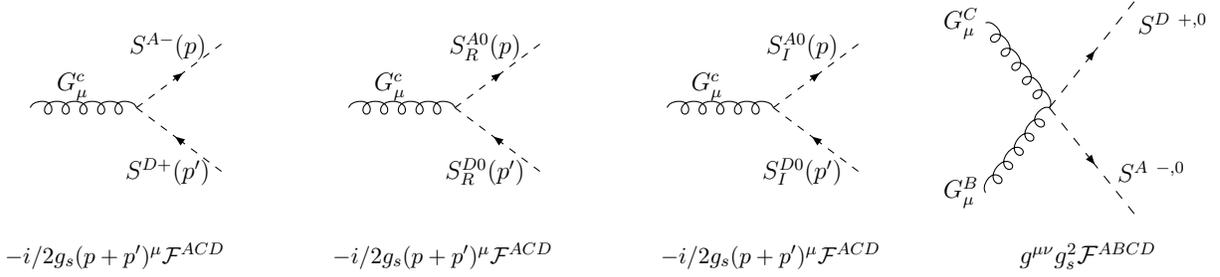}
\caption{\label{fig3} The interacting vertices of scalar octets
and gluons}
\end{center}
\end{figure}
The Feynmann rules for the interactions of scalar octets with the
SM gluons are summarized in Fig. \ref{fig3}.
In addition from the scalar potential $V(5_H,45_H)$ one gets the
following potential between the Higgs doublet $H$ and the scalar
octet $S$:%
\bea%
\nonumber V(H, S) &=& -\mu^2_H  H^\dagger H + \lambda_1 (H^\dagger
H)^2 -\mu^2_{S} S^\dagger S + \lambda_2 (S^\dagger S)^2 +
\lambda_3 (S^\dagger S) H^\dagger H + \lambda_4 S^\dagger H\
H^\dagger\ S \\ &+&\frac{1}{2}
\lambda_5\Big[ (H^\dagger\ S)^2 + (S^\dagger\ H)^2 \Big].%
\eea%
Therefore, one finds the following interacting vertices among the
$S^{\pm,0}, S^{\mp,0}$ and $h$: %
\bea%
\nonumber && h S^+ S^-: -\lambda_3 v_5 \sin \alpha
\nonumber\\
&& h S^0_{I} S^0_{I}: -(\lambda_3 + \lambda_4- \lambda_5) v_5
\sin \alpha \nonumber\\
&& h S^0_{R} S^0_{R} : -(\lambda_3 + \lambda_4+ \lambda_5) v_5
\sin \alpha. %
\eea

Finally, it is worth mentioning that after the electroweak
symmetry breaking, the octet scalars acquire the following masses
from the potential $V(45_H)$ and $V(45_H, 5_H)$: %
\bea%
m^2_{S^\pm} & = & - \mu_s^2 + \lambda^{\rm eff}_1 v_5^2 + \lambda'^{\rm eff}_1 v_{45}^2, \\
m^2_{S^0_R} & = & - \mu_s^2 + \lambda^{\rm eff}_2 v_5^2 + \lambda'^{\rm eff}_2 v_{45}^2, \\
m^2_{S^0_I} & = & - \mu_s^2 + \lambda^{\rm eff}_3 v_5^2 + \lambda'^{\rm eff}_3 v_{45}^2, %
\eea%
where $\lambda^{\rm eff}_i$ and $\lambda'^{\rm eff}_i$ are linear
combinations of the scalar couplings of $V(45_H)$ and $V(45_H,
5_H)$. Therefore, one concludes that the masses of the octet
scalars, in general, are not determined and are not universal. In
the next section, we will discuss the experimental constraints
imposed on the octet scalar masses.

%----------------------------------------------------------
\section{ Constraints on octet scalars masses}

\subsection{Constraints from $K^0-\bar{K^0}$ and
$B^0_q-\bar{B}^0_q$ mixing}

 In this section we consider possible
constraints on the mass of neutral and charged octet scalars,
$S^0$ and $S^{\pm}$, due to the experimental bounds of $\Delta
S=2$ and $\Delta B=2$ processes. The strength of $K^0 - \bar{K^0}$
mixing is described by the mass difference $\Delta M_K  = M_{K_L}
-M_{K_S}$, whose present experimental value is $\Delta M_K^{exp} =
3.483 \pm 0.006 \times 10^{-15}$ GeV, while the SM prediction is
given by $\Delta M_K^{SM}= 2.7018 \times 10^{-15}$ GeV. Therefore,
any new contribution from neutral and charged octet scalar
exchanges must be limited to the small difference between the
measured and the SM results. The lagrangian of the octet scalar
interactions with the SM
fermions can be derived from Eq.(\ref{yuk}) as%
\be%
{\cal L}_{int}^S = 2 Y_2 \left[ \bar{d}_R S^{0*} d_L + \bar{d}_R
S^- u_L \right] + Y'_4 \left[\bar{u}_R S^+ d_L - \bar{u}_R S^0 u_L
\right],%
\ee%
where $Y_2$ and $Y'_4 = 4 (Y_4 - Y_4^T)$ are generic $3 \times 3$
matrices. They contribute to the down and up quark masses as
emphasized in Eqs.(\ref{mD}),(\ref{mU}). If $M_D$ is diagonalized
by $V^d_L$ and $V^d_R$, while the $M_U$ is diagonalized by $V^u_L$
and $V^u_R$, then in the mass eigenstate basis, the couplings of
the neutral octet scalar with the down and up quarks are given by
$Y_{S^0 d_R d_L}= V_R^{d^+}. Y_2. V_{L}^d$ and $Y_{S^0 u_R
u_L}=V_R^{u^+}. Y'_4. V_{L}^u$, respectively. In minimal flavor
violation scenario \cite{light-octet}, where $Y_2 \propto Y_1$ and
$Y'_4 \propto Y'_3$, the interactions of $S^0$ with down and up
quarks become flavor diagonal and the quark couplings with charged
octet scalar depend on the quark mixing matrix $V_{CKM}$. Here we
do not adopt this assumption and consider $Y_2$ and $Y'_4$ as
generic matrices. From Eqs. (\ref{mE}), (\ref{mD}), one can
represent the Yukawa coupling $Y_2$
in terms of $M_D$ and $M_E$, namely: %
\be%
Y_2 = \frac{M_D-M_E}{8 v_{45}}, %
\ee%
which implies that
\be %
Y_{S^0 d_R d_L}= \frac{1}{4 v_{45}} \left[ M_D^{diag} - V_R^{d^+}.
M_E. V_{L}^d \right].%
\ee%
\begin{figure}[t]
\begin{center}
\includegraphics[scale=0.75]{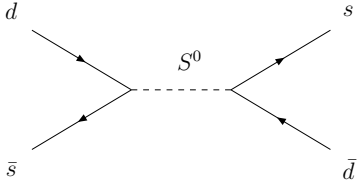}
\caption{\label{octet} Neutral octet scalar contributions to $K^0
- \bar{K^0}$ mixing}
\end{center}
\end{figure}
Due to the miss-match between the diagonalization of $M_D$ and
$M_E$, the last term generates flavor violation in the couplings
of the neutral octet scalar with down quarks, even in the basis of
diagonal charged lepton. One can assume that the quark mixing
matrix is generated mainly from the down sector, {\it i.e.}, $
V_L^d = V_{CKM}$, $V_R^d=I$, and $V_L^u = V_R^u=I$. In this case one finds%
\be %
Y_{S^0 d_R d_L}= \frac{1}{4 v_{45}} \left[ M_D^{diag} -M_E^{diag}. V_{CKM} \right].%
\ee%
In this case, the coupling
of neutral octet scalar with down and strange quarks is given by %
\be %
Y_{S^0 s_R d_L} = -\frac{m_{\mu}\lambda}{4 v_{45}} = \frac{3
m_{\mu}\lambda}{4 v \sin \beta}  ,
\ee%
while its coupling with down and bottom quarks is of order %
\be%
Y_{S^0 b_R d_L} = -\frac{m_{\tau} \lambda^3}{4 v_{45}}=
\frac{3m_{\tau} \lambda^3}{4 v \sin \beta} ,
\ee %
where $\lambda \simeq 0.21$ and $\tan \beta$ is defined as $\tan
\beta =v_2/v_1=-3 v_{45}/v_5$ In this respect, the neutral octet
scalar may contribute to the $K^0 - \bar{K^0}$ mixing at tree
level as shown in Fig. \ref{octet}, while the charged octet scalar
contribution is given by one loops similar to the SM contribution
through $W^\pm$-boson exchange. Since the mass of the charged
octet scalar $S^\pm$ is larger than the mass of the SM gauge boson
$W^\pm$, the contribution from $S^\pm$ to $K^0 - \bar{K^0}$ mixing
is typically much smaller than the SM effect. Hence no direct
constraint on the charged octet scalar mass can be imposed.

\begin{figure}[t]
\begin{center}
\epsfig{file=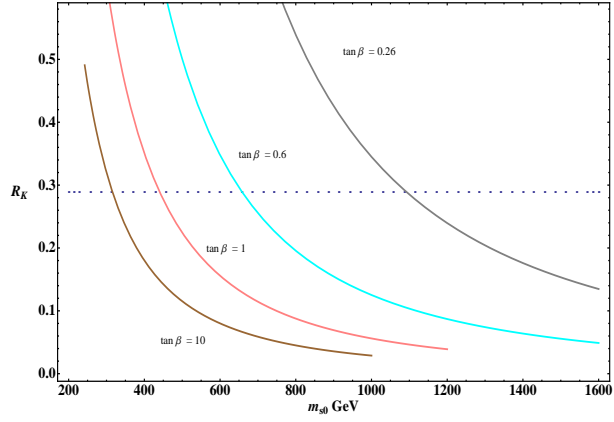,height=5.5cm,width=8.cm,angle=0}
\caption{\label{kk} The Ratio $R_K=
M_{12}^{S^0}(K)/M_{12}^{SM}(K)$ as function of the neutral octet
scalar mass $m_{S^0}$ for four values of
$tan\beta=\frac{v_2}{v_1}$. Horizontal line corresponds to the
experimental limit: $R_K=0.2891$.}
\end{center}
\end{figure}

Let us now consider the tree level contribution of neural octet
scalar $S^0$ to $\Delta S = 2$ processes, where S refers to the
strangeness quantum number. The $K_L - K_S$ mass difference
$\Delta M_K$ is defined as %
\be%
\Delta M_K = 2\mid M_{12}(K) \mid = 2\mid \langle K \mid H^{\Delta
S=2}_{eff} \mid \bar{K} \rangle  \mid ,%
\ee%
where $H^{\Delta S=2}_{eff}$ is the effective Hamiltonian for
$\Delta S = 2$ transition and the mass matrix $M_{12}(K)$ can be
written as %
\be %
M_{12}(K) =  M^{SM}_{12}(K) + M^{S^0}_{12}(K).%
\ee %
Therefore, one can write $\Delta M_K$ in the form%
\be%
\Delta M_K = \Delta M_K^{SM} \vert 1+ R_K \vert ,%
\ee%
where the ratio $R_K$ is defined as $R_K=
M_{12}^{S^0}(K)/M_{12}^{SM}(K)$. In this respect, the experimental
limit of $\Delta M_K$ \cite{Nakamura:2010zzi} leads to %
\be%
R_K \leq 0.2891. %
\ee%

The effective Hamiltonian associated to the neutral scalar
exchange is given by
\be%
H_{eff} = \sum_{i=1,2} (C_i Q_i + \tilde{C}_i \tilde{Q}_i ),%
\ee%
Where the operators $Q_i$ are given by
\be%
Q_1 = (\bar{s}_R d_L)(\bar{s}_R d_L ), \hspace{1cm} Q_2 =
(\bar{s}_R d_L)(\bar{s}_L d_R ),%
\ee%
and the Wilson coefficients $C_i$ are defined as %
\be %
C_1 = \frac{Y^2_{S^0 s_L d_R}}{m^2_{S^0}} , \hspace{1cm}  C_2 =
\frac{Y_{S^0 s_R d_L} Y_{S^0 s_L d_R} }{m^2_{S^0}}. %
\ee %
The operators $\tilde{Q}_i$ and Wilson coefficients $\tilde{C}_i$
are obtained from the $Q_i $ and $C_i$ by the exchanging $L
\leftrightarrow R$. In this case, one can easily show that
$M_{12}^{S^0}(K)$ is given by%
\be%
M_{12}^{S^0}(K) =  0.0125 \left[C_1(\mu) +
\widetilde{C}_1(\mu)\right] + 2
\times 0.017 \times C_2(\mu) %
\ee

\begin{figure}[t]
\begin{center}
\epsfig{file=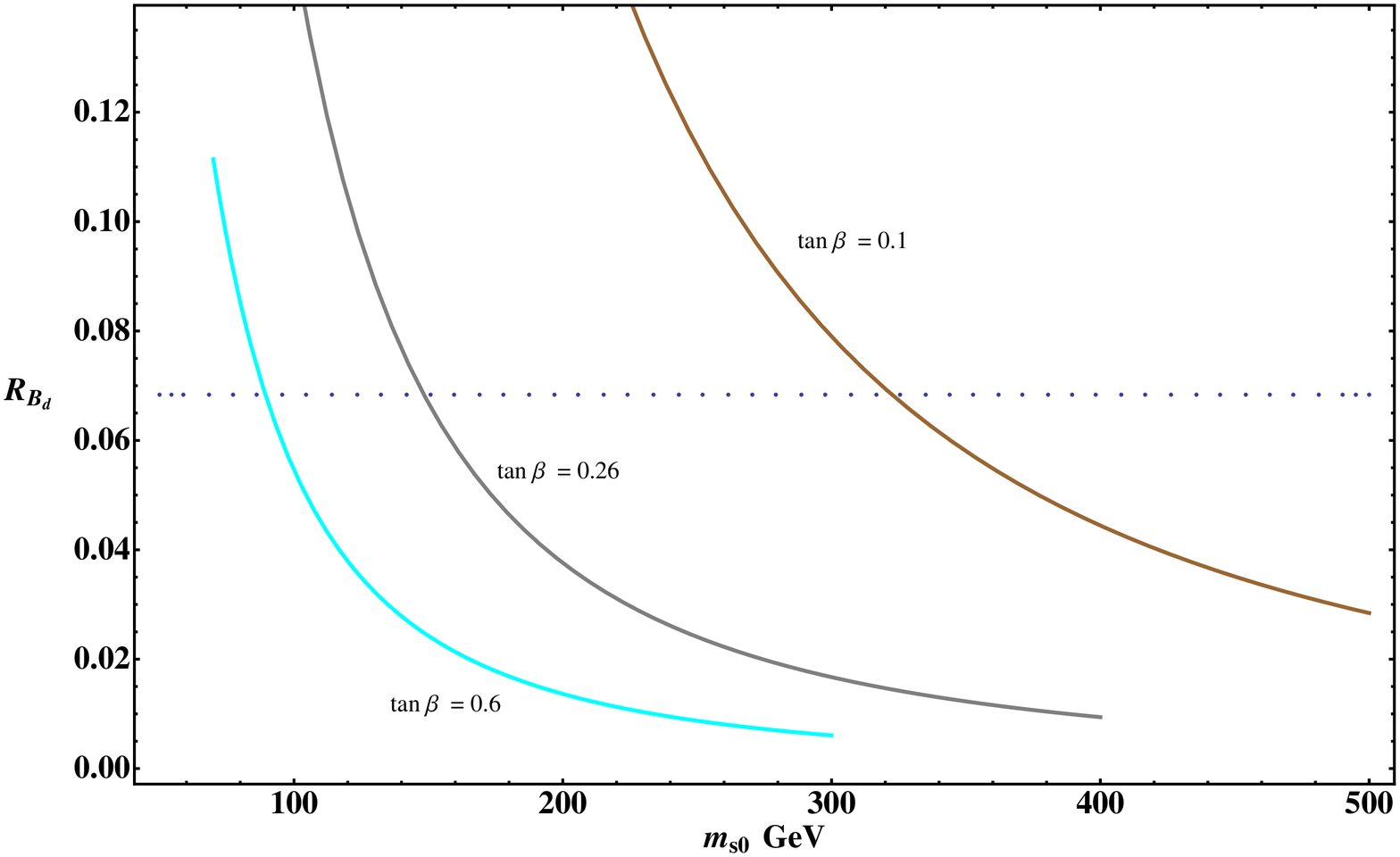,height=5.5cm,width=7.cm,angle=0}~~~~~~~~~~
\epsfig{file=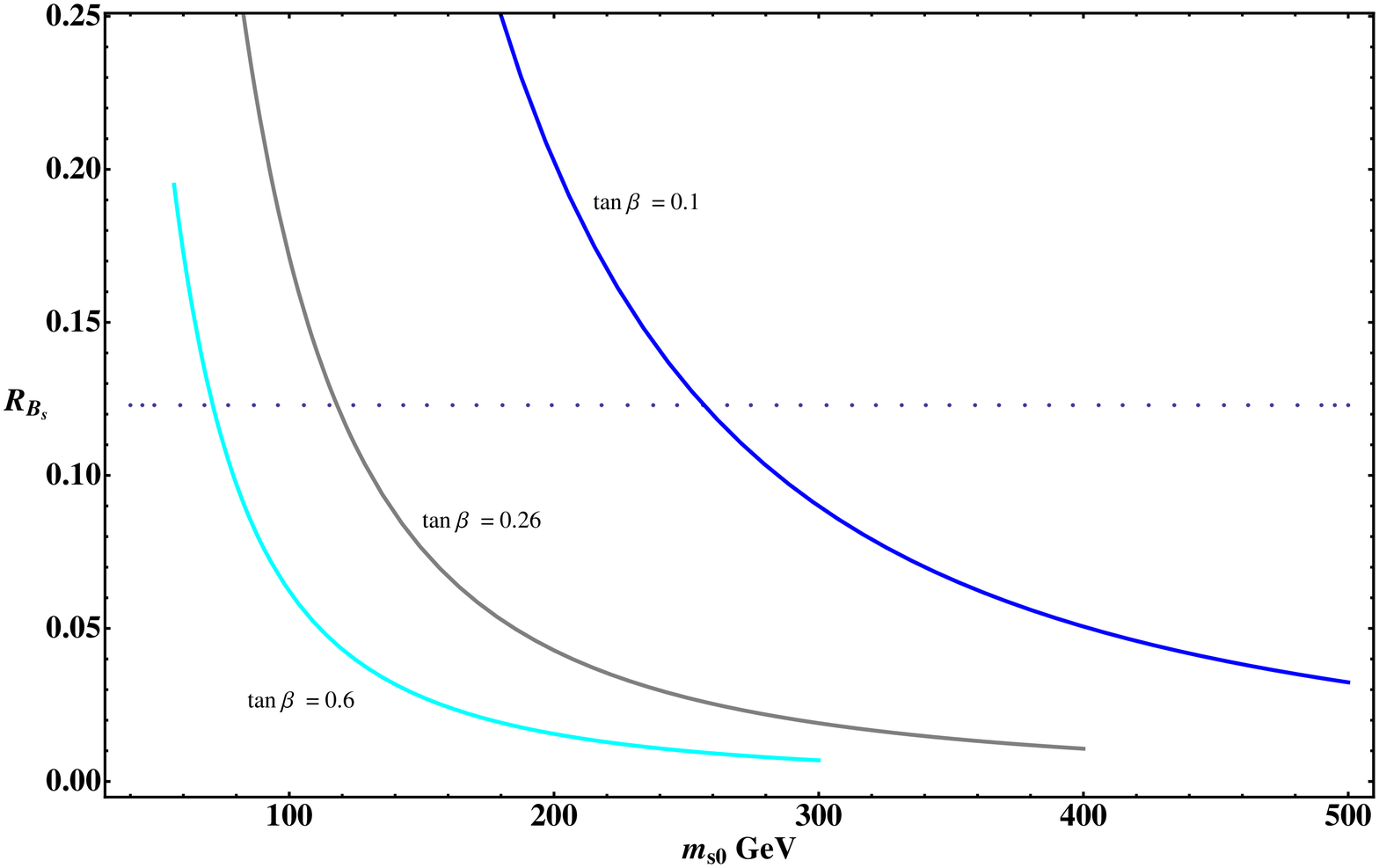,height=5.5cm,width=7.cm,angle=0}
\caption{\label{Bds} $R_{B_d}$ and $R_{B_s}$ versus $m_{S^0}$ for
$\tan\beta=0.1, 0.26, 1, 10$. }
\end{center}
\end{figure}

In Fig. \ref{kk} we show that the ratio $R_K$ as a function of the
mass $m_{S^0}$ for $\tan \beta = 0.1, 0.26, 1$, and $10$. As can
be seen from this figure, for $\tan \beta \sim {\cal O}(1)$, the
mass of neutral octet scalar can be as light as $440$ GeV. The
lower bound of $m_{S^0}$ is increased with smaller $\tan \beta$.
For instance, with $\tan \beta \sim 0.2$ one finds  that the lower
bound of $m_{S^0}$ is of order ${\cal O}(1)$ TeV.

Similarly, one can check possible constraints imposed by $B^0_q -
\bar{B}^0_q$ mixing, with $q = d, s$, on the mass of the neutral
octet scalar. The recent experimental results
\cite{Nakamura:2010zzi} for $\Delta M_{B_d}$
and $\Delta M_{B_s}$ are given by  %
\be %
\Delta M^{exp}_{B_d} = (3.337 \pm 0.033) \times 10^{-13}~ {\rm
GeV},  \hspace{1cm} \Delta M^{exp}_{B_s} = (117.0 \pm 0.8) \times
10^{-13}~ {\rm GeV}, %
\ee%
while the SM predictions are given by%
\be%
\Delta M^{SM}_{B_d} = 3.58187  \times 10^{-13}~ {\rm GeV}
\hspace{1cm} \Delta M^{SM}_{B_s} = 104.19 \times 10^{-13}~ {\rm
GeV}. %
\ee%
This implies that the ratios $R_{B_{d,s}}$ are constrained as
follows:%
\be%
\vert R_{B_{d}} \vert \leq 0.0683, \hspace{1cm} R_{B_s} \leq
0.1229. %
\ee%
These constraints on $R_{B_{d,s}}$ appear more stronger than the
constraint imposed on $R_K$, hence it may lead to more stringent
constraints on $m_{S^0}$. However, it is easy to check that the
Wilson coefficients of the $\Delta B_{d,s}=2$ are proportional to
the Yukawa couplings $Y_{bd}$ and $Y_{bs}$, which are smaller than
$Y_{sd}$. Therefore, the $S^0$ contributions to $B^0_q -
\bar{B}^0_q$ mixing are quite suppressed, hence the constraints
imposed by these process are weakened. This conclusion is
confirmed in Fig. \ref{Bds}, where we plot $R_{B_d}$ and $R_{B_s}$
versus the neutral octet scalar mass $m_{S^0}$ for $\tan \beta =
0.1, 0.26,1,10$. As can be seen from these figures that for $\tan
\beta \sim 0.6$, the $B^0_{d,s} - \bar{B}^0_{d,s}$ mixing implies
that $m_{S^0}$ can be much less than $100$ GeV.

\subsection{Direct searches constraints}

In recent years, there has been a growing interest in searching
for the octet scalar at hadron colliders, namely Tevatron and LHC
\cite{LHC-octet}. In this section we describe the recent
experimental results, which are interpreted as a lower bound on
the mass of octet scalars $(S^\pm, S^0)$.

The octet scalars can be pair-produced copiously at the LHC $gg
\to S^0 S^0$ or $gg \to S^+ S^-$. This is due to their large
couplings to gluons and their large color factors. The octet
scalars can then decay to the SM quarks without any missing
energy. Therefore, the associated signature is a pair of dijet
resonances However, the direct search of this process is very
challenging due to an enormous QCD multi-jets background that
exceeds the signal by orders of magnitude.

The production cross section of octet scalar at the LHC depends on
the mass $M_S$ and proton-proton collision energy $\sqrt{s}$.
Depend on its mass, $S$ may decay to the heaviest fermions, which
are kinematically allowed. In particular, $S^0$ decays mainly to
$t \bar{t}$ and/or $b \bar{b}$, while $S^+ \to t \bar{b}$. The
latest results with $\sqrt{s} =7$ TeV have set a $90\%$ CL limit
on the cross section of a pair of dijet that ruled out octet
scalar masses less than 287 GeV \cite{ATLAS:2012ds}. It is
important to note that the pair production cross section is almost
model independent.

\section{Octet scalar contribution to $h \to \gamma \gamma$}

As advocated in the introduction, CMS and ATLAS collaborations
observed a SM-like Higgs boson in the mass range 125-126 $GeV$
\cite{Aad:2012tfa,Chatrchyan:2012ufa}. Both collaborations
considered the following search channels: $h\rightarrow
\gamma\gamma,\ h\rightarrow ZZ^*\rightarrow 4l,\ h \rightarrow
WW^*\rightarrow 2l 2\nu$\ as well $h \rightarrow Z \gamma$, and
$h\rightarrow b\bar{b},\ \tau\bar{\tau}$. Using the full dataset
recorded by CMS and ATLAS experiments at the LHC from $pp$
collisions at center of mass energies of 7 and 8 $TeV$, both
experiments reported confirmed excess in the first three decay
channels at Higgs mass of order 125 GeV. In $h\rightarrow
\gamma\gamma$, CMS experiment observed that the signal strength,
$\sigma/\sigma_{SM}$ is given by $0.78\pm 0.27$ in case of
selected event analysis and $1.11\pm 0.31$ in the cut based
analysis \cite{moriond}. While ATLAS collaboration found that the
best fit of this signal strength is given by $1.65 \pm 0.24$
\cite{moriond}. In $ h\rightarrow ZZ^*\rightarrow 4l$, CMS
experiment measured the signal strength as $0.91^{+0.31}_{-0.24}$,
nevertheless ATLAS experiment reported an excess with signal
strength $1.5 \pm 0.4$. Finally in $ h \rightarrow WW^*\rightarrow
2l 2\nu$ both CMS and ATLAS found an excess of events above
background, which is consistent with the expectation from a SM
Higgs boson of mass $\simeq 125$ GeV.

\begin{figure}[t]
\begin{center}
\epsfig{file=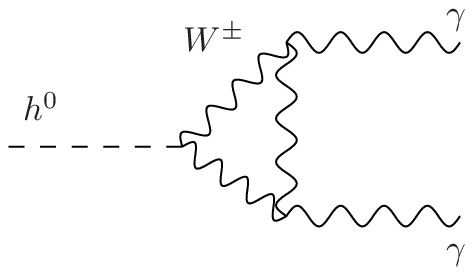,height=2.5cm,width=3.5cm,angle=0}~~~~~~~~~~~~~
\epsfig{file=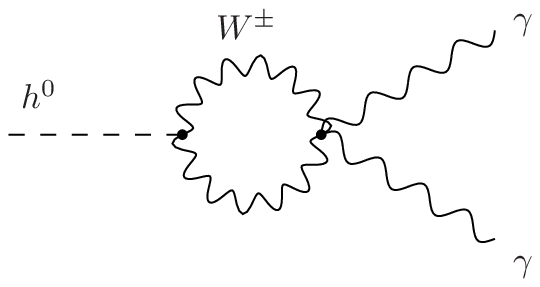,height=2.5cm,width=3.5cm,angle=0}~~~~~~~~~~~~~
\epsfig{file=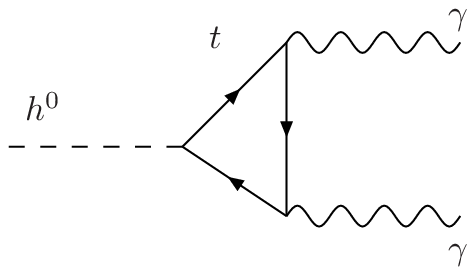,height=2.5cm,width=3.5cm,angle=0}\\
\epsfig{file=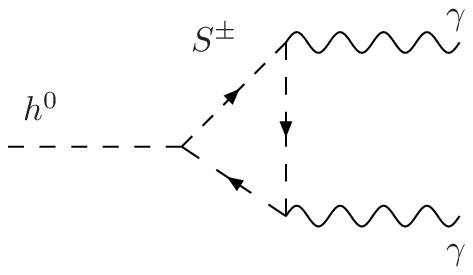,height=2.5cm,width=3.5cm,angle=0}~~~~~~~~~~~~~
\epsfig{file=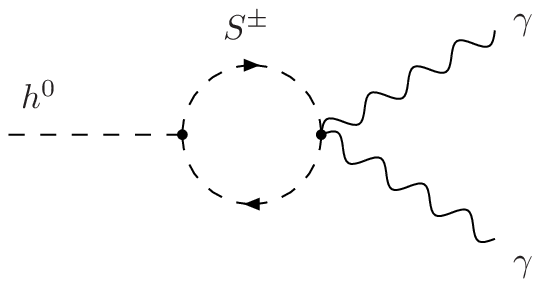,height=2.5cm,width=3.5cm,angle=0}
\caption{\label{feynman} Feynman diagrams for the decay
$h\to\gamma\gamma$ mediated by  gauge bosons $W^\pm$, top quark,
and charged octet scalars $S^\pm$.}
\end{center}
\end{figure}

It is clear that the statistical significance is not sufficient to
claim any deviation from the SM expectation. Nevertheless, the
above mentioned results indicate enhancement in the diphoton
production, with more than $2\sigma$ deviation, which could be a
very important signal for possible new physics beyond the SM.
Indeed, it has motivated many  theorists to look for possible new
physics explanation. In this section, we will emphasize that this
excess can be naturally accommodated in our low energy effective
$SU(5)$ model. In the SM, the Higgs boson decays into diphotons
through triangle loop diagram with $W^+, W^-$ and $t, \bar{t}$
exchanges. The enhancement of the diphoton decay width requires
the presence of charged particles with non-negligible coupling to
Higgs boson. In addition, this contribution of the new charged
particles should interfere constructively with the dominant SM
contribution from $W^\pm$ boson loop. As shown in previous
section, the spectrum of the low energy effective theory of
$SU(5)$ with $45_H$ contains charged color-octet scalars, $S^\pm$,
that can give a genuine contribution to the SM-like Higgs decay
into diphotons. The color-octet scalar effects on Higgs production
in gluon fusion and diphoton decay have been analyzed within
extensions of SM with color-octet scalars
\cite{Chang:2012ta,octet-Higgs}. However, our $SU(5)$ model is
very different from these phenomenological models. Therefore, we
expect to obtain new results for both Higgs decay to diphoton and
Higgs production cross section.

The contributing Feynman diagrams for the decay $h\to\gamma\gamma$
mediated by gauge bosons $W^\pm$, top quark, and $S^\pm$ are shown
in Fig. \ref{feynman}. In this case, the one-loop partial decay
width of the $H$ decay into two photons is
given by \cite{Chang:2012ta}%
\be%
\Gamma(h \to \gamma \gamma) =\frac{\alpha^2 m_H^3}{1024 \pi^3}
\Big \vert \frac{g_{hWW}}{m_W^2} Q_W^2 F_1(x_W) + N_{c,t} Q_t^2
\frac{ 2 g_{ht\bar{t}}}{m_t} F_{1/2}(x_t)  + N_{c,S}
Q_S^2\frac{g_{hSS}}{m_S^2} F_0(x_S) \Big\vert^2,
\ee%
where $x_i = m_h^2/4m_i^2$, $i=W,t,S$. The color factor and
electric charges are given by: $N_{c,t}=3$, $N_{c,S}=8$,
$Q_{W}=1$, $Q_{S}=1$, and $Q_t=2/3$. As explicitly derived in the
previous section, the Higgs couplings are given by $g_{hWW} = g
M_W \sin(\beta-\alpha)$, $g_{ht\bar{t}}= - m_t \sin \alpha/v \cos
\beta +4 Y_4 \cos \alpha$ and $g_{hS^\pm S^\mp} = - \lambda_3 v
\cos\beta \sin \alpha$.
Finally, the loop functions $F_i(x_i)$ are given by%
\bea%
F_1(x) &=& -\Big[2 x^2 + 3 x + 3 (2x -1) \arcsin^2(
\sqrt{x})\Big] x^{-2},\nonumber\\
F_{1/2}(x) &=& 2 \Big[x +(x-1) \arcsin^2(\sqrt{x})\Big] x^{-2},\nonumber\\
F_{0}(x) &=& - \Big[x -\arcsin^2(\sqrt{x})\Big] x^{-2}\nonumber.%
\eea%
\begin{figure}[t]
\begin{center}
\epsfig{file=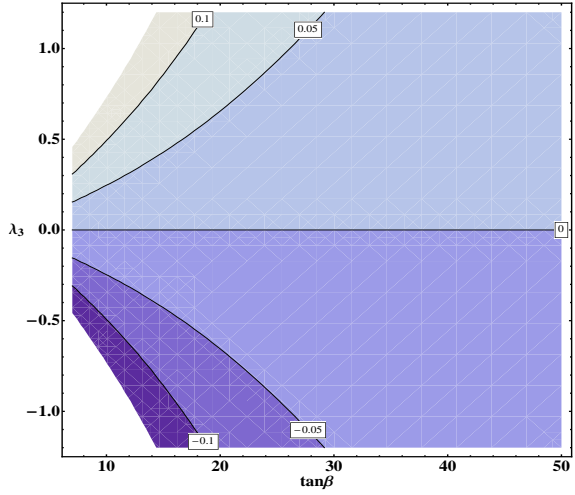,height=6.5cm,width=7.5cm,angle=0}
\caption{\label{ghSS} The coupling of SM-like Higgs with charged
scalar octet normalized by the coupling $g_{hWW}$ as a function of
$\tan \beta$ and $\lambda_3$ for $\lambda =4$. }
\end{center}
\end{figure}
For Higgs mass of order $125$ GeV and octet scalar mass of order
300 GeV, the loop functions $F_1(x_w)$, $F_{1/2}(x_t)$ and
$F_0(x_S)$ are given by $-8.32$, $+1.38$, and $0.34$ respectively.
Therefore, within the SM there is a distractive interference
between the contributions of $W$-gauge bosons and top quark. In
this respect it would be preferable to reduce the top Yukawa
coupling (specially, if it is not directly related to the top
quark mass as in our case), so that $\Gamma(h \to \gamma \gamma)$
can be enhanced. In Fig. \ref{ghww} we have shown that this can be
naturally achieved in our model and even $g_{ht{\bar{t}}}$ may
flip its sign. In this case, $\Gamma(h \to \gamma \gamma)$ becomes
much larger than the SM expectation. In addition, to allow for
constructive interference between $W$ and $S$ contributions, that
leads to an enhancement of $\Gamma(h\to \gamma \gamma)$, the
dimensionful coupling $g_{hSS}$ should be quite large and
negative. In Fig. \ref{ghSS}, we show the ratio $g_{hSS}/g_{hWW}$
as function of $\tan \beta$ and $\lambda_3$ for $\lambda =4$. As
can be seen from this figure, the coupling $g_{hSS}$ is typically
much smaller than $g_{hWW}$. It is typically less than 0.1 of
$g_{hWW}$ and it may reach $0.2$ at small $\tan \beta$. Also lower
values of $\lambda$ lead much smaller $g_{hSS}$. In this case, it
is clear that unless the charged octet scalars are very light, its
direct contribution to $\Gamma(h \to \gamma \gamma)$ is quite
marginal. Therefore, one concludes that the main effect in this
class of models is due to the reduction of the top contribution.

\begin{figure}[t]
\begin{center}
\epsfig{file=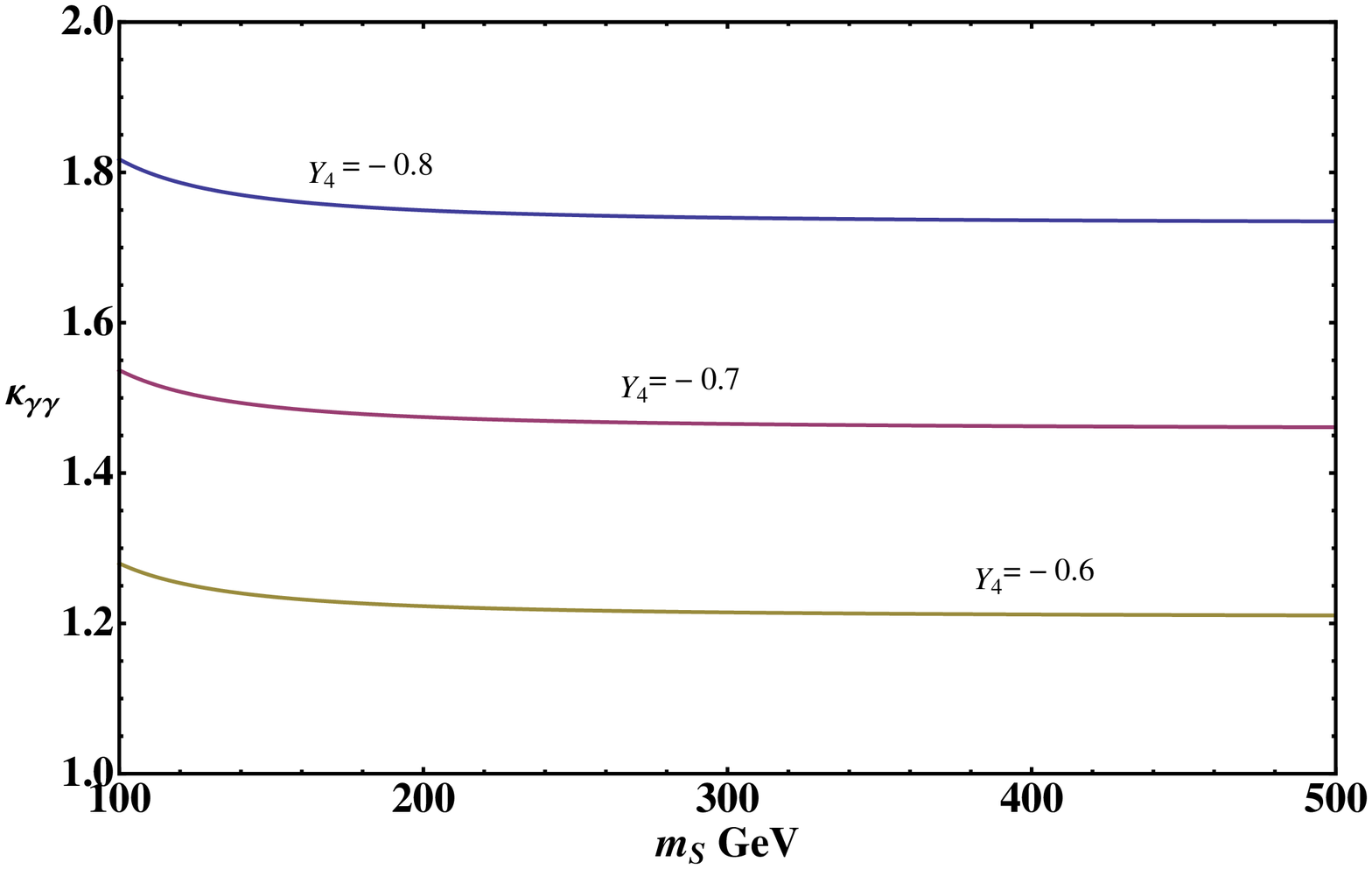,height=5.5cm,width=7.5cm,angle=0}~~~~~~
\epsfig{file=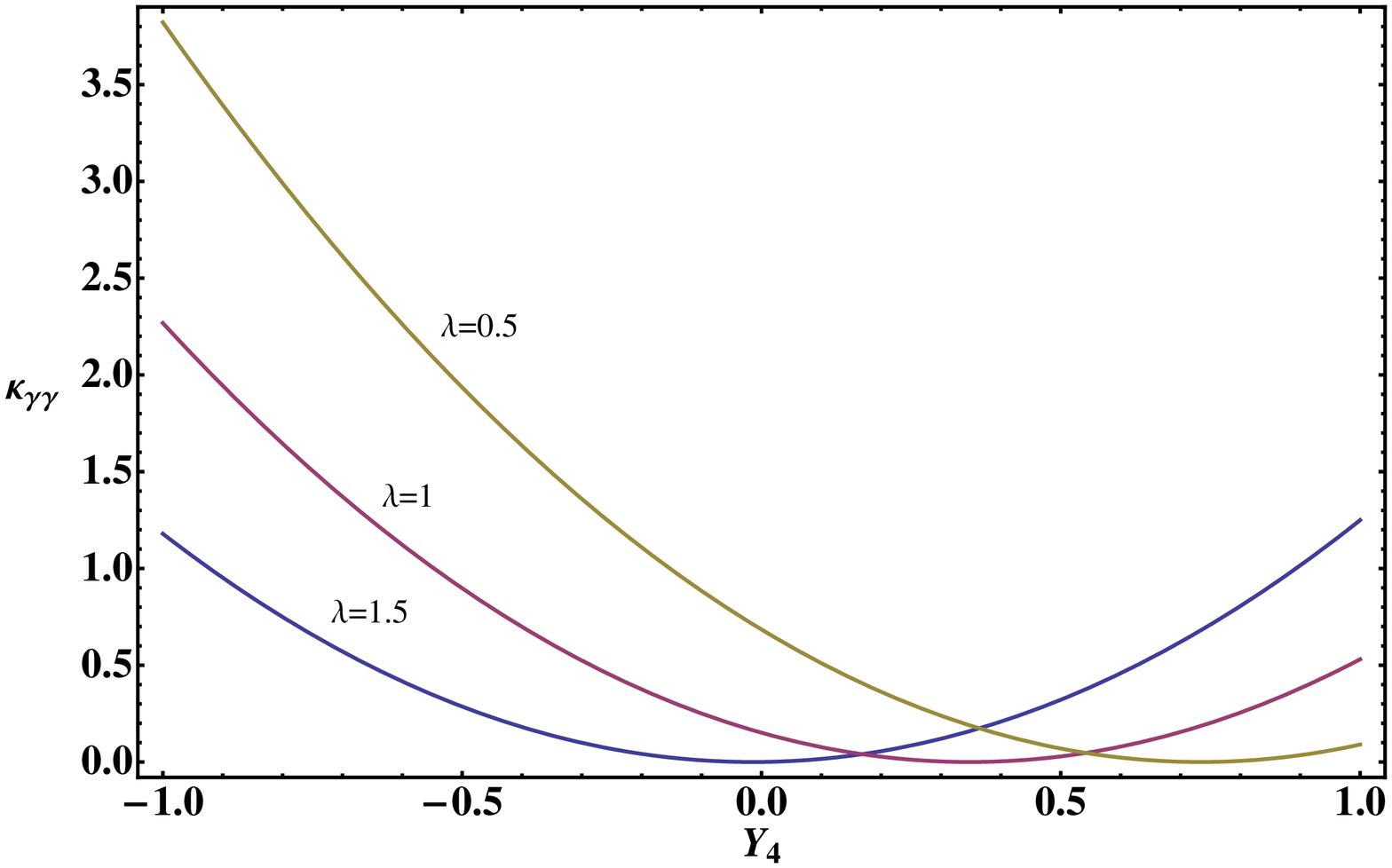,height=5.5cm,width=7.5cm,angle=0}
\caption{\label{kAA} (Left panel): The ratio $
\kappa_{\gamma\gamma} = \Gamma(h \to \gamma \gamma)/\Gamma(h\to
\gamma\gamma)^{SM}$ as a function of octet scalar mass $m_S^\pm$
for $Y_4= -0.8, -0.7,-0.6$ and $\lambda=1$, $\lambda_3=-1$, $\tan
\beta=10$. (Right panel): $\kappa_{\gamma \gamma}$ versus $Y_4$
for $\lambda = 1.5, 1, 0.5$ and $m_S^\pm = 200$ GeV,
$\lambda_3=-1$, $\tan \beta=10$.}
\end{center}
\end{figure}

In Fig. \ref{kAA} we present the ratio $\kappa_{\gamma \gamma} =
\Gamma(h \to \gamma \gamma)/\Gamma(h \to \gamma \gamma)^{SM}$ in
terms of $m_S^\pm$ for three values of $Y_4$ that induce important
suppressions on $g_{ht\bar{t}}$. As can be easily seen,
$\kappa_{\gamma \gamma}$ slightly depends on very light $m_S^\pm$.
However, $Y_4$ has a significant impact on $\kappa_{\gamma
\gamma}$. It is remarkable that for $Y_4 \simeq {\cal O}(-1)$,
$\kappa_{\gamma \gamma}$ can be of order 1.8. In addition, we
present a plot for $\kappa_{\gamma \gamma}$  versus $Y_4$ for
three different values of $\lambda$: $0.5, 1, 1.5$. From these two
plots, one can conclude that $\lambda\simeq {\cal O}(0.5)$ with
$Y_4 \simeq {\cal O}(-0.5)$ is a perfect choice in order to get
$\kappa_{\gamma \gamma} \simeq 1.6 -2$.

The Higgs signal strength of decay channel, $h \to AA$, relative
to the SM expectation is defined as
\bea%
R_{AA} &=& \frac{\sigma(pp \to h\to AA)}{\sigma(pp \to h \to
AA)^{SM}}= \frac{\sigma(pp \to h)}{\sigma(pp \to
h)^{SM}} ~\frac{BR(h\to AA)}{BR(h \to AA)^{SM}} \nonumber\\
&=& \frac{\Gamma(h \to gg)}{\Gamma(h \to gg)^{SM}}
~\frac{\Gamma_{tot}^{SM}}{\Gamma_{tot}} ~\frac{\Gamma(h \to A
A)}{\Gamma(h \to AA)^{SM}} = \kappa_{gg} .\kappa_{tot}^{-1}. \kappa_{AA} ,%
\eea%
where $\sigma(pp \to h)$ is the total Higgs production cross
section and $BR(h\to AA)$ is the branching ratio of the
corresponding channel. The total Higgs decay width is given by the
sum of the dominant Higgs partial decay widths, {\it i.e.},
$\Gamma_{tot} = \Gamma_{b\bar{b}} + \Gamma_{WW} + \Gamma_{ZZ} +
\Gamma_{\tau\bar{\tau}}$. Other partial decay widths are much
smaller and can be safely neglected. In the SM with 125 GeV Higgs
mass, these partial decay widths are given by $\Gamma_{b\bar{b}} =
2.3\times 10^{-3}$, $\Gamma_{WW}= 8.7 \times 10^{-4}$,
$\Gamma_{ZZ}= 1.1 \times 10^{-4}$, and
$\Gamma_{\tau\bar{\tau}}=2.6 \times 10^{-4}$. As shown in Fig.
\ref{ghww}, the Higgs coupling $g_{hWW}$ remains very close to the
SM value for $\lambda \simeq 1$ or at large $\tan \beta$.
Therefore, we have $\Gamma_{WW} \simeq \Gamma_{WW}^{SM}$. The
bottom Yukawa
coupling in our model takes the form%
\be %
Y_{hb\bar{b}} \simeq - \frac{3 m_b + m_\tau}{4 v \cos \beta} \sin
\alpha ,%
\ee%
which can be of order the SM Yukawa coupling $Y_b^{SM} = m_b/v$ if
$\sin \alpha \sim \frac{4}{3} \cos \beta$. This condition can be
satisfied if $\lambda < 1$. We will adopt this constraint in our
analysis so that $\Gamma_{b\bar{b}}$ remains intact and hence
$\Gamma_{tot} \simeq \Gamma_{tot}^{SM}$.

\begin{figure}[t]
\begin{center}
\epsfig{file=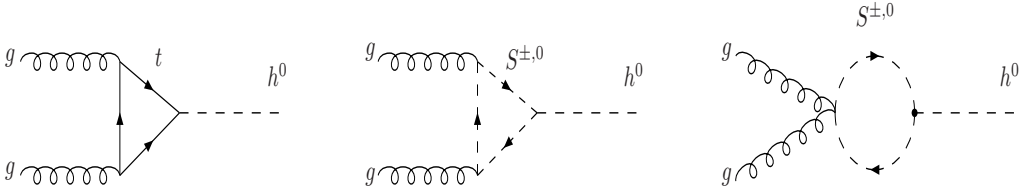,height=2.5cm,width=13.5cm,angle=0}
\caption{\label{fusion} Gluon fusion $ gg \to h$ in $SU(5)$
effective model, mediated by top quark, and charged and neutral
color-octet scalars.}
\end{center}
\end{figure}

\begin{figure}[t]
\begin{center}
\epsfig{file=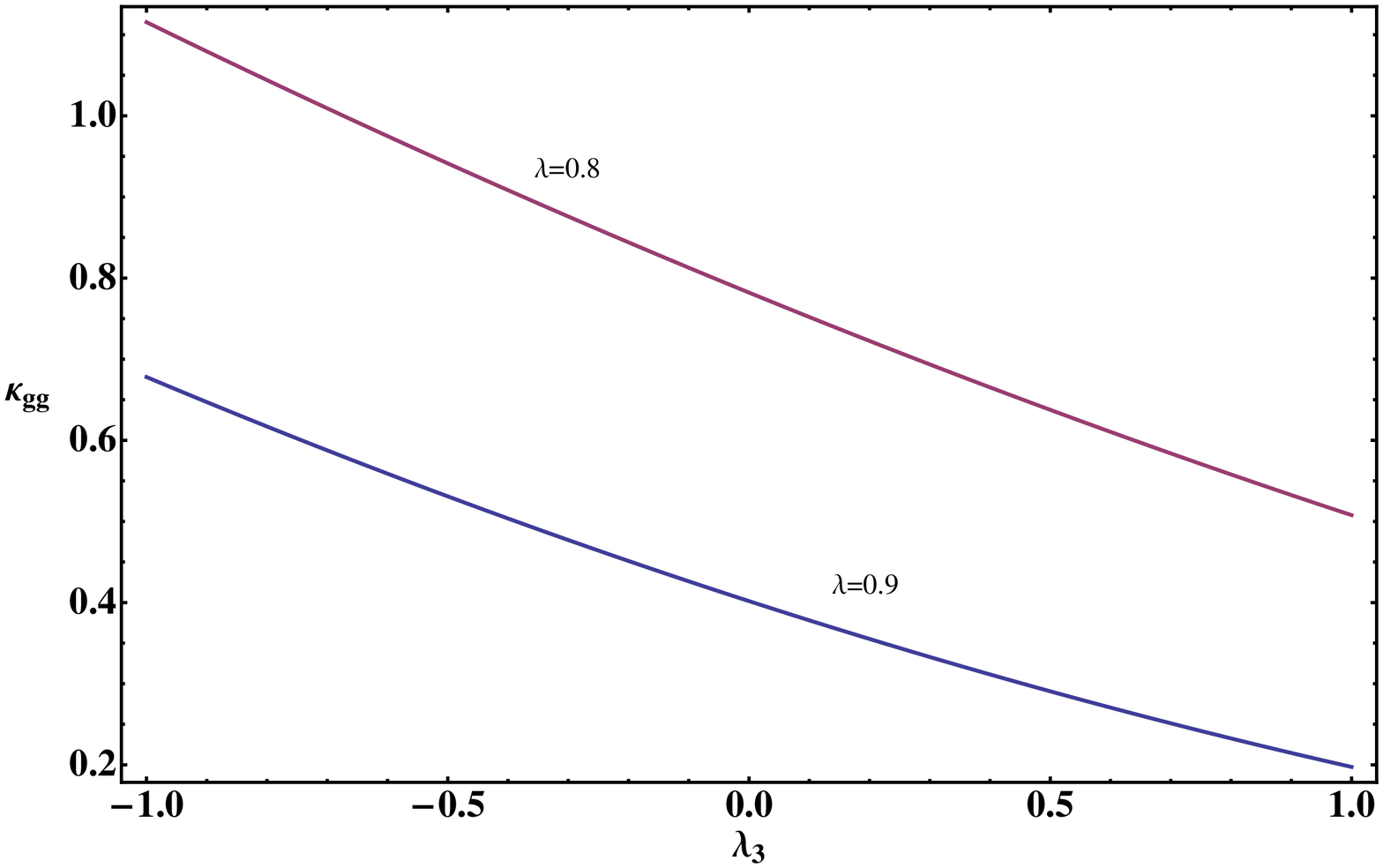,height=5.5cm,width=7.5cm,angle=0}~~~~~~
\epsfig{file=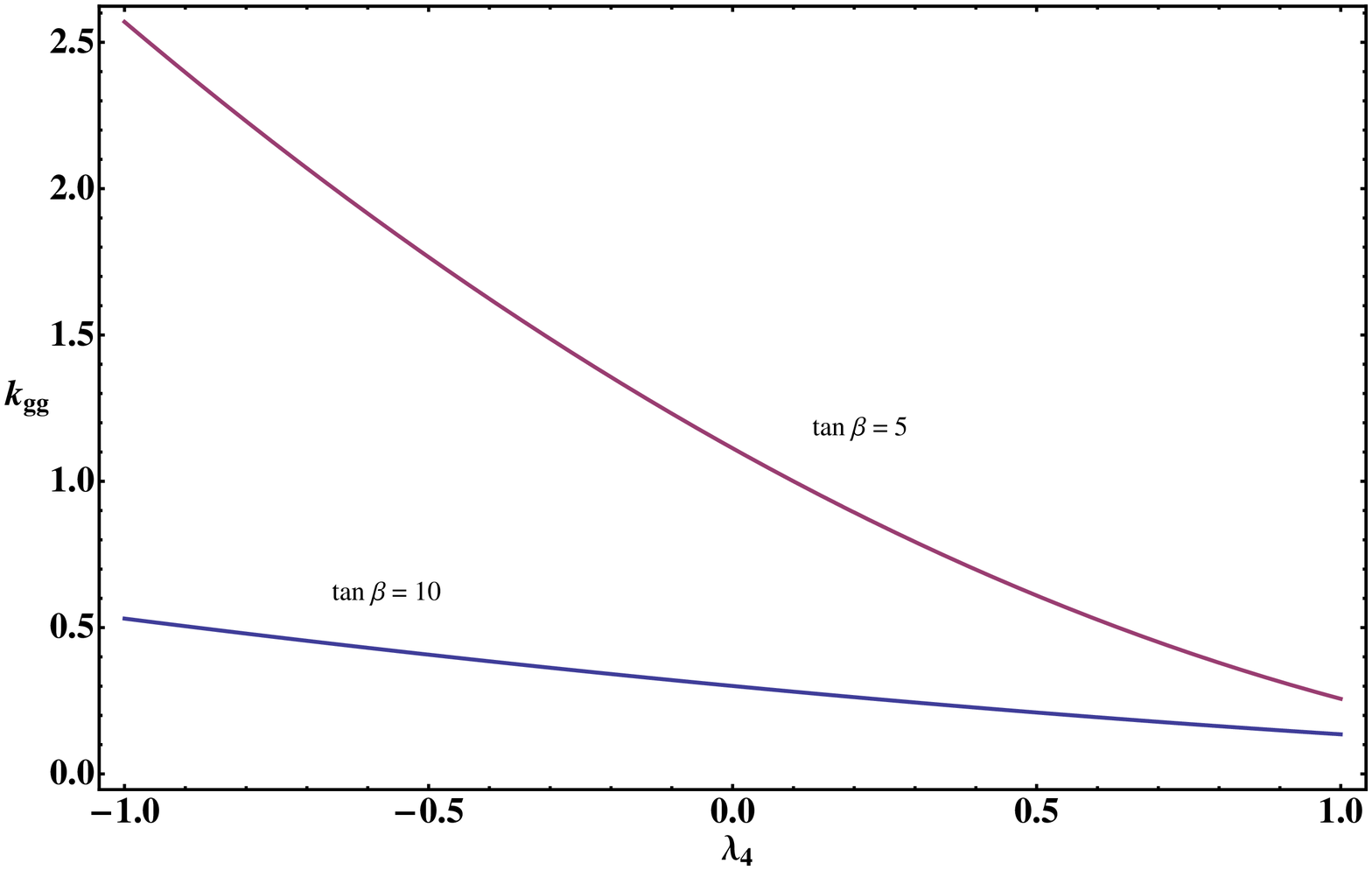,height=5.5cm,width=7.5cm,angle=0}
\caption{\label{kgg} (Left panel): The ratio $ \kappa_{gg} =
\frac{\Gamma(h \to g g)}{\Gamma(h\to gg)^{SM}}$ as a function of
the coupling $\lambda_3$ for $\lambda= 0.8, 0.9$ and $m_S=300$
GeV, $Y_4=-0.8$, $\lambda_4=-1$, $\tan \beta=10$. (Right panel):
$\kappa_{g g}$ versus $\lambda_4$ for $\tan \beta = 5, 10$ and
$m_S=300$ GeV,$\lambda=0.9$, $Y_4=-0.8$, $\lambda_3=-0.5$.}
\end{center}
\end{figure}

Now we turn to the Higgs production cross section in our $SU(5)$
effective model. At the LHC the dominant process for the Higgs
production is the gluon-gluon fusion. In the SM the gluon fusion
mechanism is mediated by top-quark via one loop triangle diagram.
However, in our model the gluon fusion for the SM-like Higgs can
be also obtained through the exchange of neutral and charged
color-octet scalars,  as shown in Fig.\ref{fusion}. The lowest
order cross section can be written as
\be%
\hat{\sigma}_{Lo}(gg\rightarrow h)= \frac{\pi^2}{8m_h}
\Gamma_{LO} (h \rightarrow gg) \delta(\hat{s}-m_h^2).%
\ee%
where $\hat{s}$ is the center of mass energy and
$\delta(\hat{s}-m_h^2)$ is the Breit-Wigner form of the Higgs
boson width, which is given by
$$\delta(\hat{s}-m_h^2) = \frac{1}{\pi} \frac{\hat{s}
\Gamma_h/m_h}{(\hat{s}-m_h^2)^2 + (\hat{s} \Gamma_h/m_h)^2}.$$
The partial decay width $\Gamma(h \to gg)$ is given by \cite{Chang:2012ta}%
\be %
\Gamma(h \rightarrow gg) =\frac{\alpha_s^2 m_h^3}{128 \pi^3}
\Big|C(r_t) \frac{2 g_{ht\bar{t}}}{m_t}   F_{1/2}(x_t) + C(r_S)
\frac{g_{hS^\pm S^\mp}}{m^2_{S^\pm}} F_{0}(x_{S^\pm})+
C(r_S) \frac{g_{hS^0S^0}}{m^2_{S^0}} F_{0}(x_{S^0}) \Big|^2, %
\ee%
where $C(r)$ is the $SU(3)$ representation index, which is defined
as ${\rm Tr}[T^a_r T^b_r] = C(r) \delta^{ab}$ with $C(3)=C(r_t)
=1/2$ and $C(8)=C(r_s)=3$.
In the above expression, it was assumed that $m_{S^0_R}=
m_{S^0_I}$. Therefore, the coupling $g_{hS^0S^0}$ is given by
$$g_{hS^0S^0} = - (\lambda_3+ 2 \lambda_4) v \cos \beta \cos\alpha.$$
Thus, large values of $\lambda_{3,4}$ and small values of $\tan
\beta$ are preferred to enhance the $g_{hS^0S^0}$ coupling. In
Fig. \ref{kgg} we present the ratio $\kappa_{gg} = \Gamma(h\to
gg)/\Gamma(h\to gg)^{SM}$ versus $\lambda_3$ for  $\lambda= 0.8,
0.9$ and versus $\lambda_4$ for $\tan \beta = 5, 10$. As can be
seen from these plots that the neutral octet-scalar can give a
significant contribution to $\Gamma(h \to gg)$, so that it
compensates the suppressions caused by: $(i)$ the reduction of top
Yukawa coupling, $(ii)$ the negative effect of charged octet
scalar. In this case, the region $0.8 \lsim k_{gg} \lsim 1$, which
is preferred by best fit analysis of the recent experimental
results \cite{Buckley:2012em}, is quite accessible.

\begin{figure}[t]
\begin{center}
\epsfig{file=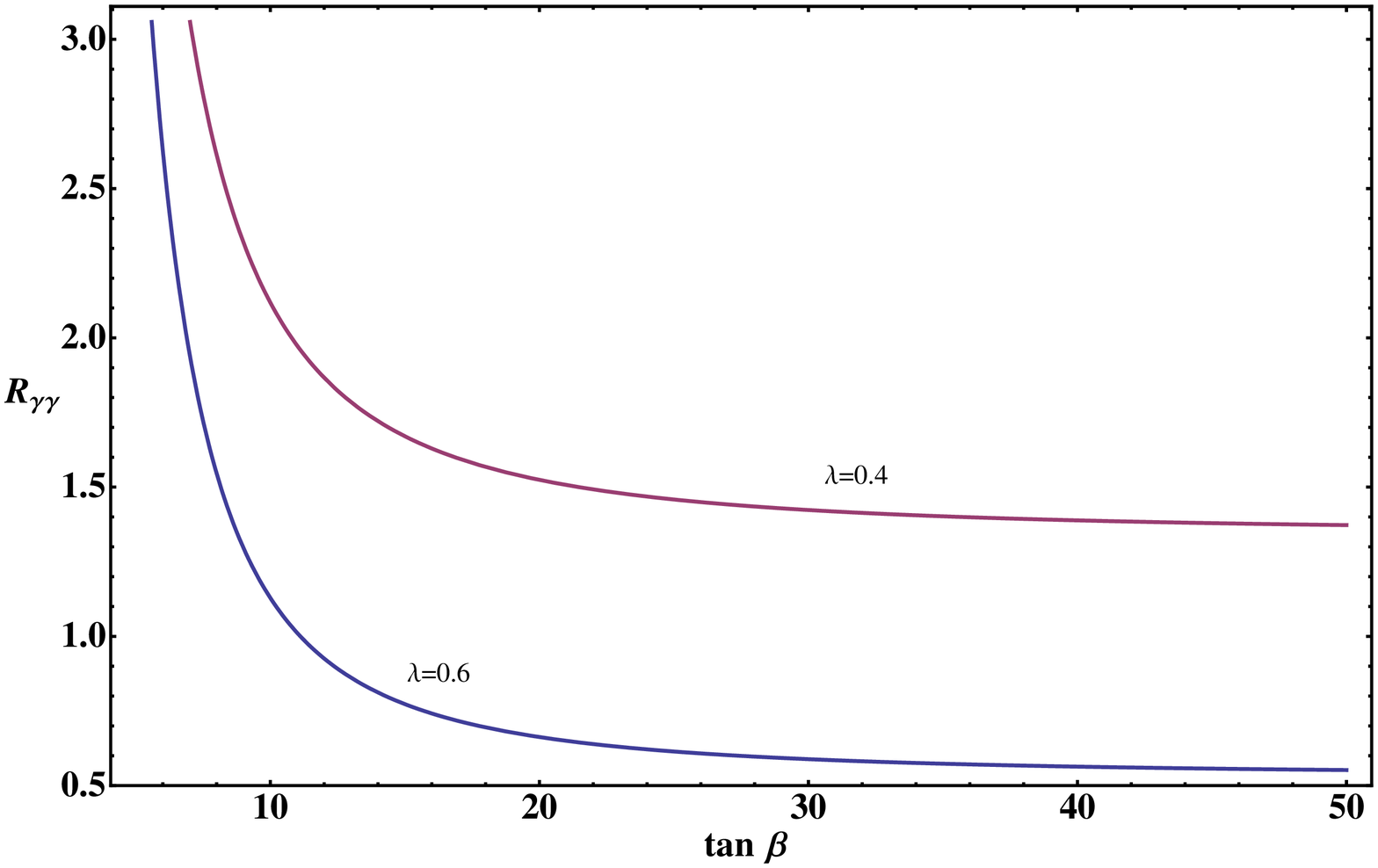,height=5.5cm,width=7.5cm,angle=0}~~~~~~
\epsfig{file=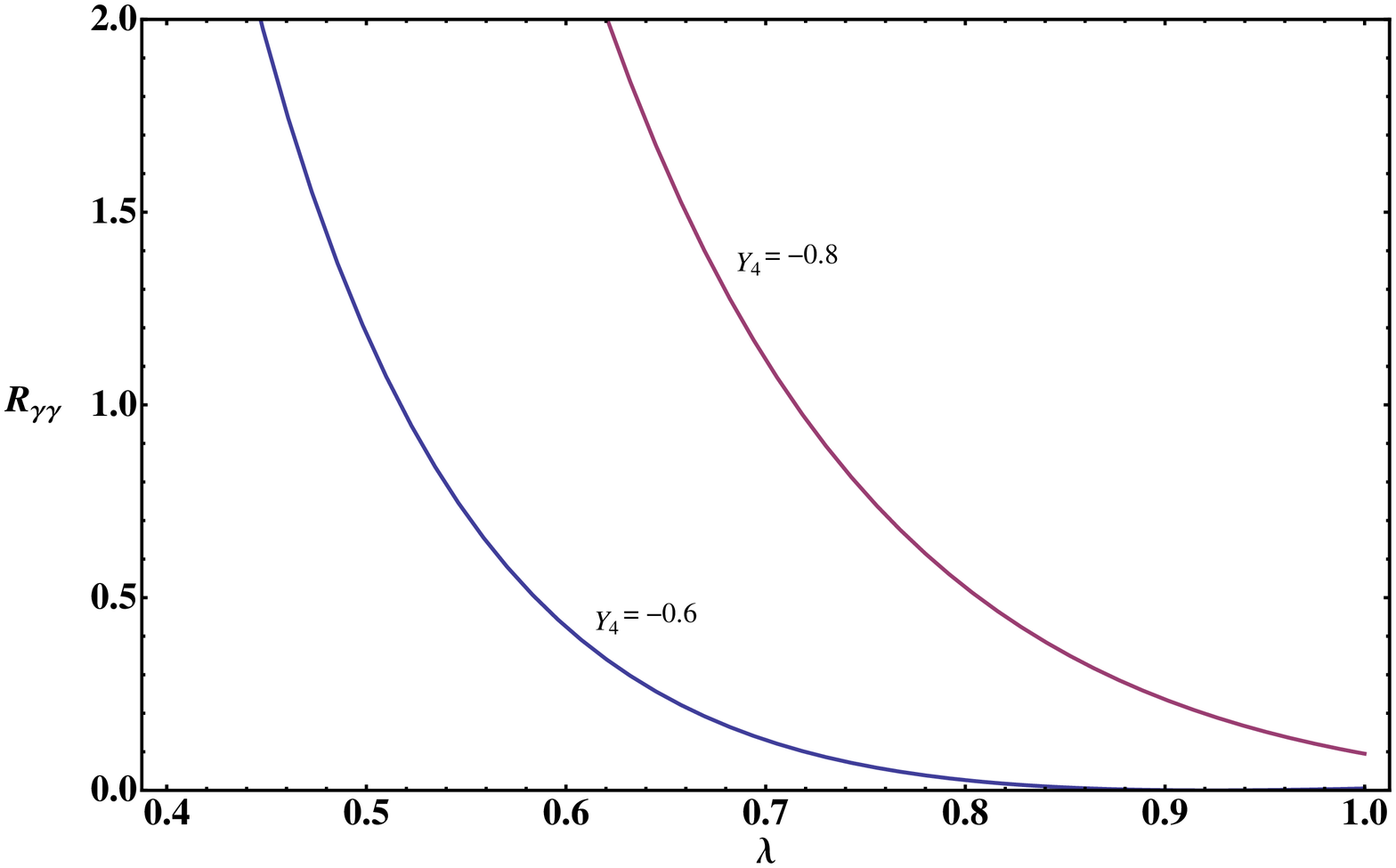,height=5.5cm,width=7.5cm,angle=0}
\caption{\label{signal} (Left panel): The signal strength $
R_{\gamma \gamma}$ as a function of $\tan \beta$ for $\lambda=
0.4, 0.6$ and $m_S=300$ GeV, $Y_4=-0.7$, $\lambda_3=\lambda_4=-1$.
(Right panel): $R_{\gamma \gamma}$ versus $\lambda$ for $Y_4=-0.6,
-0.8$ and $m_S=300$ GeV, $\lambda_3=\lambda_4=-1$, $\tan \beta
=10$.}
\end{center}
\end{figure}

From the results of $\kappa_{\gamma\gamma}$ and $\kappa_{gg}$ with
$\Gamma_{tot} \simeq \Gamma_{tot}^{SM}$, one can easily see that
the recent experimental measurement of signal strength $R_{\gamma
\gamma}$ by ATLAS and CMS collaboration can be easily accommodated
in our model. The sign of color octet-scalar couplings can be
fixed based on the final results of $R_{\gamma \gamma}$. If it is
confirmed that $R_{\gamma \gamma} > 1$, then the coupling
$\lambda_3$ should be of order ${\cal O}(-1)$ and $Y_4 \simeq
{\cal O}(-0.5)$ so that the contribution of top quark is reduced.
On the other hand, if $R_{\gamma \gamma}$ is proven to be less
than one as indicated by the latest result of CMS experiment, then
$\lambda_{3}$ should be positive and $Y_4$ should quite small so
that the top quark effect remains as in the SM or even bigger. In
Fig. \ref{signal} we display the signal strength $ R_{\gamma
\gamma}$ as a function of $\tan \beta$ for $\lambda= 0.4, 0.6$ and
universal octet scalar mass $m_S=300$ GeV, $Y_4=-0.7$,
$\lambda_3=\lambda_4=-1$. Also we plot $R_{\gamma \gamma}$ versus
$\lambda$ for $Y_4=-0.6, -0.8$ and $m_S=300$ GeV,
$\lambda_3=\lambda_4=-1$, $\tan \beta =10$.

\section{Conclusions}
In this paper we have derived the low energy effective model of
$SU(5)$ grand unified field theory with extending the Higgs sector
by $45_H$-plet. We showed that this model is an extension of the
SM with another Higgs doublet and color-Octet scalar doublet. We
analyzed the flavor violation constraint of the octet-scalar
masses. We found that the $K^0-\bar{K}^0$ mixing impose a
stringent bound on the neutral octet scalar mass if $\tan \beta <
1$. We have also studied all possible contributions to the light
neutral Higgs decay into diphoton. We emphasized that the charged
octet scalars may provide a constructive interference with the SM
$W^\pm$ gauge bosons effects, which enables an enhancement of the
branching ratio of $h \to \gamma \gamma$. However, it turns out
that the most significant impact on the diphoton decay width in
this model is due to a possible suppression for top-Yukawa
coupling with SM-like Higgs or even flipping its sign that leads
to important enhancement in $\Gamma(h \to \gamma \gamma)$ that
accounts for the measured signal strength. In addition, we have
studied the impact of the neutral octet-scalar on the gluon fusion
Higgs production cross section. We showed that with this
contribution one can keep $\kappa_{gg} \sim {\cal O}(1)$, while
$\kappa_{\gamma \gamma}\sim {\cal O}(1.6)$. So the apparent
tension between enhancement of diphoton decay rate and suppression
of $\sigma(gg \to h)$ is resolved in this class of models.

%

%------------------------------------------------------------------------------

\section*{Acknowledgements}
SK thanks The Leverhulme Trust (London, UK) for financial support
in the form of a Visiting Professorship to the University of
Southampton. SM is financed in part through the NExT Institute.

%----------

\end{document}